\documentclass[]{aa} 
\usepackage{graphicx} 
\usepackage{rotating,subfigure,amssymb,afterpage} 
\usepackage{txfonts} 
\usepackage{natbib} 
\usepackage[pdftitle={XMM Cluster Structure Survey}, 
pdfauthor={H. B\"ohringer et al.}, 
pdfsubject={}, pdfkeywords={}, bookmarks, bookmarksopen,  
colorlinks=true, linkcolor=blue, citecolor=blue, anchorcolor=blue, 
urlcolor=blue]{hyperref}  
\newcommand{\propsim}{\lower 3pt \hbox{$\, \buildrel {\textstyle 
      \propto}\over {\textstyle \sim}\,$}} 
\begin{document} 
   \title{The Representative XMM-Newton Cluster Structure Survey ({\sl 
       REXCESS}) of an X-ray Luminosity Selected Galaxy Cluster Sample}

   \author{H. B\"ohringer\inst{1}, P. Schuecker\inst{1}, 
     G.W. Pratt\inst{1}, M. Arnaud\inst{2}, T.J. Ponman\inst{3},  
    J.H. Croston\inst{4}, S. Borgani\inst{5}, R.G. Bower\inst{6}, 
    U.G. Briel\inst{1}, C.A. Collins\inst{7},  
    M. Donahue\inst{8}, W.R. Forman\inst{9}, A. Finoguenov\inst{1}, 
    M.J. Geller\inst{9}, L. Guzzo\inst{10},  
    J.P. Henry\inst{11}, R. Kneissl\inst{12}, J.J. Mohr\inst{13}, 
    K. Matsushita\inst{14}, C.R. Mullis\inst{15}, T. Ohashi\inst{16},  
    K. Pedersen\inst{17}, D. Pierini\inst{1}, H. Quintana\inst{18}, 
    S. Raychaudhury\inst{3},   
    T.H. Reiprich\inst{19}, A.K. Romer\inst{20}, P. Rosati\inst{21},  
    K. Sabirli\inst{22} R.F. Temple\inst{3}, P.T.P. Viana\inst{23,24}, 
    A. Vikhlinin\inst{9}, G.M. Voit\inst{8}, Y.-Y. Zhang\inst{1} 
          } 
 
   \offprints{H. B\"ohringer} 
 
   \institute{$^1$ Max-Planck-Institut f\"ur extraterrestrische Physik, 
                 D 85748 Garching, Germany, {\tt hxb@mpe.mpg.de}\\ 
              $^2$ Service d'Astrophysique, CEA Saclay, L'Ormes des 
              Merisiers, F-91191 Gif-sur-Yvette Cedex, France\\  
              $^3$ School of Physics and Astronomy, University of 
              Birmingham, Edgbaston, Birmingham B15 2TT, UK\\  
              $^4$ School of Physics, Astronomy and Mathematics, 
              University of Hertfordshire,   
                      College Lane, Hatfield AL10 9AB, UK\\ 
              $^5$ Dipartimento di Astronomia dell'Universit\'a di 
              Trieste, via Tiepolo 11, I-34133 Trieste, Italy\\  
              $^6$ Institute for Computational Cosmology, Department 
              of Physics, University of Durham, South Road, Durham DH1 
              3LE, UK\\  
              $^7$ Liverpool John Moores University, Liverpool,U.K.\\ 
              $^8$ Department of Physics and Astronomy, Michigan State 
              University, BPS Building, East Lansing, MI 48824, USA \\ 
              $^9$ Harvard Smithsonian Center for Astrophysics, 60 
              Garden Street, Cambridge MA, USA\\  
              $^{10}$ INAF - Osservatorio Astronomico di Brera-Merate, 
              via Bianchi 46, I-23807 Merate, Italy\\  
              $^{11}$ Institute for Astronomy, University of Hawai'i, 
              2680 Woodlawn Drive, Honolulu, HI 96822, USA \\  
              $^{12}$ Max-Planck-Institut f\"{u}r Radioastronomie, Auf dem
              H\"{u}gel 69, D-53121 Bonn, Germany\\
              $^{13}$ Department of Physics, University of Illinois, 
              1110 West Green Street, Urbana, IL 61801, USA\\
              $^{14}$Department of Physics, Tokyo University of Science, 
               1-3 Kagurazaka, Shinjyuku-ku, Tokyo 162-8601, Japan\\  
              $^{15}$ Department of Astronomy, University of Michigan, 
              918 Dennison, 500 Church Street, Ann Arbor, MI 48109, 
              USA \\  
              $^{16}$ Department of Physics, Tokyo Metropolitan 
              University, Hachioji, Tokyo 192-0397, Japan\\  
              $^{17}$ Dark Cosmology Centre, Niels Bohr Institute, 
              University of Copenhagen, Juliane Maries Vej 30, DK-2100 
              Copenhagen, Denmark\\  
              $^{18}$ Departamento de Astronomi\'a y Astrofisica, 
              Pontificia Universidad Catolica de Chile, Casilla 306, 
              Santiago 22, Chile\\  
              $^{19}$ Argelander Institute for Astronomy (AIfA), Bonn 
              University, Auf dem H\"ugel 71, D-53121 Bonn, Germany\\  
              $^{20}$ Astronomy Centre, University of Sussex, Falmer, 
              Brighton BN1 9QJ, UK\\  
              $^{21}$ European Southern Observatory, 
              Karl-Schwarzschild-Strasse 2, D-85748 Garching, 
              Germany\\  
              $^{22}$ Department of Physics, Carnegie Mellon 
              University, 5000 Forbes Avenue, Pittsburgh, PA 15217\\  
              $^{23}$ Departamento de Matem\'{a}tica Aplicada da 
              Faculdade de Ci\^{e}ncias da Universidade do Porto, Rua 
              do Campo Alegre 687,  
                     4169-007 Porto, Portugal\\ 
              $^{24}$ Centro de Astrof\'{i}sica da Universidade do 
              Porto, Rua das Estrelas, 4150-762 Porto, Portugal\\   
             } 
 
   \date{Received 13 November 2006 / Accepted 5 March 2007}

\abstract 
{The largest uncertainty for cosmological studies using clusters of galaxies  
is introduced by our limited knowledge of the statistics of galaxy cluster 
structure, and of the scaling relations between observables and 
cluster mass.}   
{To improve on this situation we have started an XMM-Newton Large 
Programme for the in-depth study of a representative sample of 33 
galaxy clusters, selected in the redshift range $z= 0.055$ to 0.183 
from the REFLEX Cluster Survey, having X-ray luminosities above 
$0.4\times 10^{44} h_{70}^{-2}$ erg s$^{-1}$ in the 0.1 - 2.4 keV 
band. This paper introduces the sample, compiles properties of the 
clusters, and provides detailed information on the sample selection 
function.}  
{We describe the selection of a nearby galaxy cluster sample that makes 
optimal use of the XMM-Newton field-of-view, and provides nearly 
homogeneous X-ray luminosity coverage for the full range from poor 
clusters to the most massive objects in the Universe.} 
{For the clusters in the sample, X-ray fluxes are derived and compared 
to the previously obtained fluxes from the ROSAT All-Sky Survey.  We 
find that the fluxes and the flux errors have been 
reliably determined in the ROSAT All-Sky Survey analysis used for the 
REFLEX Survey. We use the sample selection function documented in 
detail in this paper to determine the X-ray luminosity function, and 
compare it with the luminosity function of the entire REFLEX sample. 
We also discuss morphological peculiarities of some of the sample 
members.} 
{The sample and some of the background data given in this introductory 
paper will be important for the application of these data in the 
detailed studies of cluster structure, to appear in forthcoming 
publications.} 
\authorrunning{B\"ohringer et al.} 
\titlerunning{The XMM Cluster Structure Survey (REXCESS) } 
   \maketitle 
%
 
\section{Introduction} 
 
Galaxy clusters, as the largest well-defined dark matter haloes, are 
fundamental probes for the evolution of the cosmic large-scale 
structure. Furthermore, they are ideal astrophysical laboratories for 
the study of numerous aspects of cosmic evolution, such as galaxy and 
star formation histories. The two major elements of such studies are 
(i) putting further constraints on cosmological models (e.g. Haiman et 
al. 2005, "white paper"), and (ii) the study of the evolution of the 
galaxies, both in the thermal and chemical imprint of the galaxies on 
the intracluster medium (e.g. Voit 2005, Borgani et al. 2005), and in 
the effect of the cluster environment on the formation and evolution 
of the galaxies (e.g. Croton et al. 2006).  
 
With the recent progress in observational cosmology, a concordance 
cosmological model has become established, requiring two so far 
unknown ingredients -- dark matter and dark energy (e.g. Perlmutter et 
al. 1999, Schmidt et al. 1999, Schuecker et al. 2003b, Spergel et 
al. 2006, 2003, Tegmark et al. 2006). Since the expansion of the 
Universe and the growth of structure depends very sensitively on both 
dark constituents, a detailed study of large-scale structure evolution 
in the recent past ($z \sim 2 $ to $ 0$), by means of galaxy clusters, 
can provide new insights into the nature of the dark components and 
provide tighter constraints on cosmological models. Galaxy clusters 
were among the first probes to be used to constrain dark energy 
models (Wang \& Steinhardt 1998), and their importance as cosmological 
probes is increasingly being recognised (e.g. Rosati et al. 2002, 
Schuecker et al., 2003a,b, Majumdar \& Mohr 2004, Allen et al. 2004, 
Haiman et al. 2005).  
 
Accurate mass estimates of the surveyed galaxy clusters are a 
prerequisite for such cosmological applications. X-ray observations 
are still the most attractive method to detect and characterize galaxy 
clusters. Not only is X-ray selection an approximate selection by 
cluster mass, due to the tight X-ray luminosity mass relation 
(Reiprich \& B\"ohringer 2002, Reiprich 2006), but it also provides a 
zeroth-order mass estimate through observables like the X-ray 
luminosity or X-ray temperature. For cosmological applications we need to have 
a precise knowledge of both the applied observable-mass relation and its 
intrinsic scatter. While such relations have been investigated for 
specially selected clusters (e.g. Ettori et al. 2004, Arnaud et 
al. 2005, Vikhlinin et al. 2006, Zhang et al. 2006, 2007, Pedersen \& 
Dahle 2006), accurate calibrations for such scaling relations 
for a representative, unbiased sample of X-ray flux selected galaxy 
clusters at different epochs are still needed. Thus, a major 
goal of the present project is to provide a calibration baseline of 
representative scaling relations for the cluster population in the 
nearby Universe.  
 
Quite apart cosmological applications, the form of the scaling 
relations between observables and mass, and the relations among 
different observables, provide important insights into the structure 
of galaxy clusters and the thermal structure of their intracluster 
medium. To first order, the scaling relations of observable cluster 
parameters as a function of cluster mass can be understood as 
self-similar, and have been successfully numerically simulated by 
purely gravitational structure growth (e.g. Bryan \& Norman 1998). 
However, second order effects, best observed in the low mass systems, 
show an altering of these simple self-similar relations as a 
consequence of energy input from other sources, related to galaxy and 
star formation processes and the cooling of dense intracluster gas 
regions (e.g. Kaiser 1991, Bower 1997, Ponman et al. 1999, 2003, Voit 
\& Bryan 2000, Voit et al 2003, Pratt \& Arnaud 2005, Pratt et 
al. 2006). Therefore a detailed study of the scaling relations, in 
combination with the study of the enrichment of the intracluster 
medium by heavy elements, in conjunction with numerical modeling, 
provides important insights into the cosmic history of star formation 
and the processes that govern galaxy evolution (e.g.  Pearce et 
al. 2000, Borgani et al. 2001, 2004, Muanwong et al. 2002, Kay et 
al. 2004, Finoguenov et al. 2003). 
 
For both tests of cosmological models and studies of structure growth, 
a precise knowledge of the scaling relations, and well-measured 
cluster masses for large and systematically constructed cluster 
samples are the most important prerequisites. XMM-Newton, with its 
high sensitivity and the possibility of spatially resolved 
spectroscopy, provides the best means to approach this fundamental 
task. An inspection of the observational data on galaxy clusters in 
the XMM-Newton archive shows that it is impossible to construct a 
representative, statistically unbiased X-ray luminosity-selected 
sample, optimized for the XMM field-of-view, from the existing 
observations.  Therefore we have successfully requested observing time 
for a comprehensive survey of X-ray structure of a representative 
cluster sample involving 33 galaxy clusters (where 3 data sets have 
been retrieved from the XMM-Newton archive). 
 
A systematic study of clusters is absolutely necessary.  There is for 
example a clear difference in the properties of clusters selected for 
their regularity, and those selected from flux or luminosity criteria, 
which include a wide range of morphologies. The scaling relations of 
X-ray luminosity and mass that apply to regular or compact cooling 
core clusters are expected to be different from those of dynamically 
less evolved clusters. This has been demonstrated for the X-ray 
luminosity - mass/temperature relation, showing that high central 
surface brightness objects, or objects described as classical cooling 
flows, have a significantly higher normalization than other clusters 
(O'Hara et al. 2006, Chen et al. 2006). 
 
Thus, relations obtained for symmetric, apparently relaxed clusters 
will not be applicable to general cluster surveys. Therefore our 
primary goal is the calibration of the scaling relations for a 
statistical sample of clusters, selected by X-ray luminosity alone 
(the criterion most commonly used in cosmological applications of 
clusters). We also hope to establish the present data as a benchmark 
sample for studies in other wavelengths.  In this context, contrary to 
most previous studies where researchers would choose the most regular 
clusters for an intercomparison, we want to provide a special 
incentive to also observe and reconstruct the more complex, apparently 
unrelaxed objects with different techniques of structure and mass 
measurements. 
 
Therefore, some of the major goals of this project are to better 
characterize and understand (i) the relations of observables such as 
X-ray luminosity, temperature, and characteristic radius with cluster 
mass, (ii) the source of the scatter in these relations, (iii) the 
dynamical states of the clusters via inspection of temperature, 
entropy and pressure maps as well as by the comparison of X-ray and 
optical spectroscopic observations (guided by simulations), (iv) the 
statistics of cluster mergers and the frequency of cluster cooling 
cores as a function of cluster mass; both cosmologically very 
important diagnostics (e.g. Schuecker et al. 2001b), (v) entropy 
profiles of the clusters as probes of the thermal and star formation 
history in the clusters, (vi) metal abundances in clusters as a 
function of various observational parameters, and (vii) the variation 
of the cluster mass and mass-to-light ratio profiles. 
 
   \begin{figure} 
   \begin{center} 
   \includegraphics[width=\columnwidth]{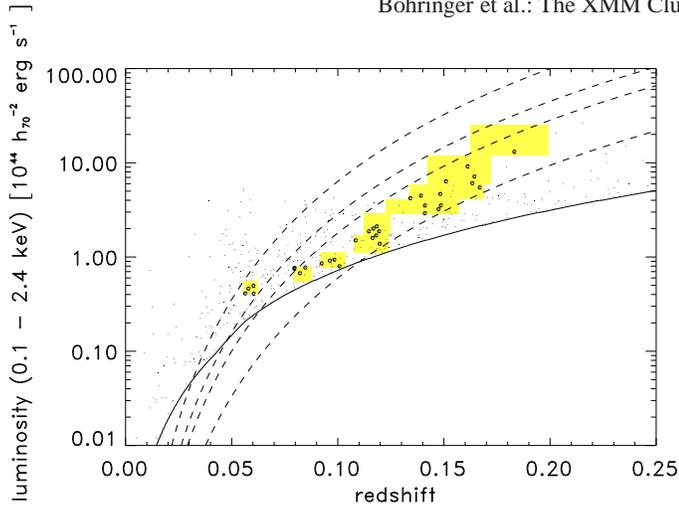} 
      \caption{X-ray luminosity-redshift distribution 
  of the REFLEX sample (small dots: entire REFLEX sample including   
  clusters with less than 30 cts and $N_H > 6 \times 10^{20}$ cm$^{-2}$),  
  and the representative subsample (encircled dots) 
  selected from the regions marked by colored boxes. The solid line indicates the survey  
  flux limit. The dashed lines show the distances at which $r_{500}$ is 7, 9,  
  10, and 12 arcmin (from right to left), for given X-ray luminosity, respectively.}  
         \label{Fig1} 
  \end{center} 
   \end{figure} 

We have also started a series of projects for observations of these 
clusters at other wavelengths, such as multicolour-photometry with the 
wide-field imaging camera, WFI, at the 2.2m MPG/ESO telescope, 
spectroscopic observations at Magellan, Sunyaev-Zeldovich observations 
with APEX, radio observations with GMRT (Giant Metrewave Radio 
Telescope), and high resolution X-ray observations with Chandra. A 
series of structure studies of these clusters is close to publication: 
analysis of the X-ray surface brightness and gas density distribution 
(Croston et al., for the novel analysis method see Croston et 
al. 2006), two-dimensional projected density, temperature, 
pseudo-pressure and pseudo-entropy maps (Finoguenov et al., similar to 
the analysis in Finoguenov et al. 2005, 2006), substructure analysis 
by a center shift method (Temple et al.), comparison of cluster 
structure of these observations with simulated clusters by means of a 
power ratio method (B\"ohringer et al.), and one study on the cluster 
temperature profiles, showing a high degree of universality of these 
profiles (outside the central regions) has been published (Pratt et 
al. 2007). 
 
There is other work in progress to obtain similar information on 
representative cluster samples at other epochs, with which the present 
work will be combined in the future. An almost complete set of 
observations with XMM-Newton and Chandra has been performed on the 
brightest 63 galaxy clusters in the sky away from galactic plane, 
the HIFLUGCS sample (Reiprich \& B\"ohringer 2002), providing an 
account at even lower redshifts (although with a somewhat less ideal 
field-of-view criterion). The data reduction of 
this sample is in progress (Hudsen et al., and Reiprich et al., 
2006). The REFLEX-DXL sample (Zhang et al. 2006) is a 
representative sample of the most X-ray luminous clusters in the 
redshift interval $z = 0.27$ to $0.31$.  After the launch of our 
project another XMM-Newton large programme was granted for the 
systematic study of cluster structure at intermediate redshifts of 
$z \sim 0.3 - 0.6$ (P.I. M. Arnaud). For even larger redshifts a 
systematic study of the RDCS clusters from Rosati et al. (1998, 2002, 
e.g. Ettori et al. 2004) and of the 160 deg$^2$ of Vikhlinin et 
al. (1998) is ongoing (e.g. Kotov \& Vikhlinin 2006).Together these 
studies shall provide a comprehensive view of the  
evolution of the cluster structure, and the corresponding scaling 
relations. 
 
The derivation of the results presented in this paper is based on the 
``concordance cosmological model'' with values of the normalized 
densities of $\Omega_m=0.3$, $\Omega_{\Lambda} = 0.7$ and a Hubble 
constant of $H_0 = h_{70}~~ 70$ km s$^{-1}$ Mpc$^{-1}$, if not 
explicitly stated otherwise.

\section{Sample construction} 
 
\subsection{Primary considerations} 
 
For the construction of an unbiased, X-ray selected cluster sample we 
use as a parent sample the REFLEX survey catalogue, which is presently 
the largest, well controlled cluster catalogue (B\"ohringer et 
al. 2004). The quality of the sample has been demonstrated by showing 
that it can provide reliable measures of the large-scale structure 
without distorting artifacts (Collins et al. 2000, Schuecker et 
al. 2001a, Kerscher et al. 2003), yielding cosmological parameters in 
good agreement within the measurement uncertainties with the 3year 
WMAP results (Schuecker et al.  2003a, b, Stanek et al. 2006, Spergel 
et al. 2006; note that this good agreement with the new WMAP data is 
also true for other cluster studies e.g. Voevodkin \& Vikhlinin 2004, 
Henry 2004).  Moreover, the study of the galaxy cluster number density 
and the measured large-scale clustering provide consistent 
cosmological results. 
 
REFLEX, based on the ROSAT All-Sky Survey (Tr\"umper 1993), is a 
highly complete ($> 90\%$) flux limited ($F_X [0.1 - 2.4 {\rm keV}] 
\ge 3 \cdot 10^{-12}$ erg s$^{-1}$ cm$^{-2}$) cluster sample, covering 
4.24 ster in the southern extragalactic sky ($\delta \le 2.5 \deg $, 
$|b_{II}| \ge 20\deg $, with regions covered by the Magellanic clouds 
excluded - B\"ohringer et al. 2001). The variation of the sky coverage 
as a function of flux is small and is well documented in the REFLEX 
catalogue paper (B\"ohringer et al. 2004). There is a residual risk 
that a substantial part of the X-ray emission detected for these 
clusters comes from AGN in the cluster or in the background. We have 
estimated, however, that the fraction of clusters with severe 
contamination by AGN emission is smaller than 9\%. 
 
The basic criteria for the selection of the present subsample are the 
following:

$\bullet$ We restrict the redshifts to $z\le 0.2$ to obtain a census 
of the local Universe. 
 
$\bullet$ The basic selection criterion is X-ray luminosity, with no 
preference for any particular morphological type. Thus the sample 
should be representative of any local, high quality, unbiased X-ray 
survey, a survey of the type applicable to cosmological model testing. 
 
$\bullet$ To best assess the scaling relations, the selection has been 
designed to provide a close to homogenous coverage of the X-ray 
luminosity range. The chosen luminosity regime, $L_X = 0.407 - 
20 \times 10^{44} h_{50}^{-2}$ erg s$^{-1}$ in the 0.1-2.4 keV 
rest frame band\footnote{Originally selected as $L_X = 0.75 - 32 \times 10^{44} 
h_{50}^{-2}$ erg s$^{-1}$ for an Einstein-De Sitter Universe}, provides 
clusters with estimated temperatures above 2 keV. Thus the spectrum 
of selected objects covers the range from poor systems to the most 
massive clusters. Lower temperature systems, groups of galaxies, are 
excluded because their study requires a larger observational effort 
than the handful of additional data points that can be afforded here. 
 
$\bullet$ We aim for a good global characterization of the clusters, 
and thus wish to detect cluster emission out to the fiducial radius 
$r_{500}$, the radius inside which the mean cluster mass density is 
500 times the critical density of the Universe.  This has been shown 
by simulations to provide one of the best measures of the size of the 
virialized dark matter system (Evrard et al. 1996). 
 
$\bullet$ The distances of the objects are selected to optimally use the 
field-of-view, angular resolution, and photon collection power of the  
{\sl XMM-Newton} observatory. For the data reduction we use 
the region of the target fields outside about 10 - 11 arcmin to assess the  
X-ray background of the observation. This is to enable a  
comparison of the properties of the target background and the 
background field to correct for background variations.   
 
$\bullet$ We use well defined selection criteria such that the space density 
of the sample and any subset of it is well defined by the selection function. 
 
These selection requirements cannot be met by a simple flux-limit cut. 
In particular, to meet the condition of a nearly homogeneous 
luminosity coverage, we decided to draw the sample from the 
luminosity-redshift distribution in 8 luminosity bins containing a 
similar number of clusters. The FoV criterion then calls for a 
staircase like distribution of these bins in the $L_X$-redshift 
diagram shown in Fig. 1 (Each bin is almost volume limited with small 
corrections explained at the end of Section 2.2). To obtain 
sufficient statistics, the minimum number of clusters in such a sample 
is of the order of 30. The affordable amount of XMM-Newton observing 
time for deep enough studies of a cluster does not allow for a much 
larger number of targets. Therefore we decided to plan for the 
selection of four clusters per luminosity bin. 
 
\subsection{Sample construction method} 
 
We start the selection by choosing 9 luminosity bins of nearly equal 
logarithmic width, as defined in Table 1 and Fig. 1, with the 
calculation of the  
redshift for which the most luminous cluster in the bin has an 
apparent radius $r_{500}$ in the sky of 9 arcmin. This radius is 
calculated by means of the X-ray luminosity-temperature relation taken 
from Ikebe et al. (2002) \footnote{All formulae are given for a 
concordance cosmology model as specified in section 1, where $h_{100} 
= H_0 / 100$ km s$^{-1}$ Mpc$^{-1}$}: 
 
\begin{equation} 
{ L_X \over 10^{44} {\rm erg~s}^{-1}} = 0.02 \left( {T_x \over {\rm 1~keV}} 
\right) ^{2.5} h_{100}^{-2}~~~~. 
\end{equation} 
 
We do not apply a redshift evolution correction here, since the 
HIFLUGCS sample studied by Ikebe et al. (2002) has a very similar 
redshift distribution as a function of luminosity to the present 
sample, and is therefore directly applicable. With the estimated 
temperature, $r_{500}$ can then be derived by means of the temperature 
- radius relation from Arnaud et al. (2005): 
 
\begin{equation} 
r_{500} = 0.773~ {\rm Mpc}~~ h_{100}^{-1} E(z)^{-1} \left({T_X \over 5{\rm keV}}\right)^{0.57}  
\end{equation} 
 
$$~= 0.753~{\rm Mpc}~~ h_{100}^{-0.544} E(z)^{-1} \left({ L_X \over 10^{44} {\rm erg~s}^{-1}}\right) 
^{0.228}$$  
 
with $E(z) = h(z)/h_0$. The second-lowest dashed line in Fig. 1 
corresponds to the relation of X-ray luminosity and redshift for 
which $r_{500}$ appears as 9 arcmin. Then we select the bins as 
follows: 
 
\begin{itemize} 
\item{(1) The upper left corner of bins 5, 6 and 7 are defined by the 
    9 arcmin radius line, effectively fixing the lower redshift 
    boundary. We collect the 4 clusters we wish to have in the bin by 
    increasing the redshift. The outer redshift boundary of the bin is 
    defined by the midpoint between the last cluster in the sample and 
    the first cluster outside.}  
 
\item{(2) For practical reasons we have not strictly applied this rule 
    to all bins. Using this criterion, luminosity bins 2, 3 and 4 
    extend very close to the nominal flux limit (the solid curve in 
    Fig. 1). For these bins we decided to start filling the  bins from 
    the high redshift side, touching the flux limit with the lower 
    right corner of the bin, and filling the bin by collecting 
    clusters at lower redshifts. The inner boundary of the redshift 
    bin is defined by the midpoint in redshift between the last 
    cluster in the bin and the first cluster at lower redshift 
    outside.}   
 
\item{(3) The lowest luminosity objects (bin 1) have a lower surface 
    brightness, and we do not expect very much emission at large 
    radii. To better use the field-of-view of XMM-Newton, and to 
    increase the flux from these clusters, we moved this lowest 
    luminosity bin to the limit where the most luminous cluster would 
    have an apparent $r_{500}$ of 12 arcmin.}  
 
\item{(4) For the most luminous clusters (bins 8 and 9), which are 
    very rare, we increased the search volume at low redshift by 
    allowing the most luminous cluster to have an $r_{500}$ of 10 
    arcmin. We also relax the interstellar column density constraints 
    and allow values of $N_H$ larger than $6 \times 10^{20}$ 
    cm$^{-2}$.}  
 
\item{(5) In bin 9, we find only one cluster in the region between an 
    $r_{500}$ of 10 arcmin and a redshift of 0.2. This cluster is 
    A1689, which has already been observed with XMM-Newton and the data 
    are available in the archive. The outer redshift boundary of this 
    bin is again determined by the midpoint to the next 
    object at higher redshift.}  
\end{itemize} 
 
The original sample was constructed from a preliminary REFLEX 
catalogue. Between the first complete catalogue construction and the 
final revision and subsequent publication of the catalogue in 
B\"ohringer et al. (2004), a series of new galaxy redshifts became 
available in the literature, publicly available data bases, and 
through our own observations. This led to improved cluster redshifts. 
Since the redshift boundaries of our relatively small cluster sample 
are very tight, there was an unavoidable scatter of objects across the 
boundaries.  We checked the typical changes in the redshifts of the 
REFLEX sample that cause the scattering of the clusters in redshift 
space, and found that about 10 - 15\% of the clusters experienced 
shifts of 400 - 2000 km s$^{-1}$, resulting in the above described 
effect.  
 
This has led us to reconstruct the selection scheme by applying the 
same criteria such that the originally-selected clusters are still 
contained in the survey volume. The mid point rule for the outer or 
inner redshift boundary gives us the flexibility to reconstruct using 
the new redshifts. The price paid is that a small number of new 
clusters appear in the bins, which then have to be corrected for in 
the selection function. With this revision we also changed the 
luminosity values from a critical density universe to a concordance 
cosmological system ($\Omega_m = 0.3$, $\Omega_{\Lambda}=0.7$, and 
$h_{70} = 1$), which results in the inclusion of two additional 
clusters inside the bins. The advantage of this reconstruction is that 
the new survey selection function is fully compatible and reproducible 
with the published REFLEX data set. Fig. 1 provides an account for the 
complete selection scheme. The resultant redshift boundaries are 
listed in Table 1. The total number of clusters ending up in the bins 
is given in column 7 of this Table, and the extra clusters are 
explicitly listed in Table 4.

\begin{table*} 
      \caption{The luminosity-redshift bins used for the selection of the sample (for $H_0=70$ km/s/Mpc, 
                $\Omega_m = 0.3$, $\Omega_{\Lambda}=0.7$)} 
         \label{tab5} 
      \[ 
         \begin{array}{lllllllllll} 
            \hline 
            \noalign{\smallskip} 
 {\rm Bin~no.} & N_{Cl} & L_X (min) & L_X (max) & z_{min} & z_{max} & N_{tot} & {\bf vol. cov.} & sky cov. & density^c & density^d \\ 
    &     &     &     &     &       &   & correction^a & correction^b &  &  \\ 
(1) & (2) & (3) & (4) & (5) & (6) & (7) & (8) & (9) & (10) & (11)\\ 
            \noalign{\smallskip} 
            \hline 
            \noalign{\smallskip} 
1  & 4& 0.407 & 0.55 & 0.0555 & 0.06215  & 4  & 0.94 & 0.775 & 5.28(-6) & 4.95(-6) \\ 
2  & 4& 0.55  & 0.78 & 0.0794 & 0.0877   & 4  & 0.86 & 0.775 & 3.33(-6) & 3.12(-6) \\ 
3  & 4& 0.78  & 1.13 & 0.0920 & 0.1037   & 7  & 0.87 & 0.775 & 8.72(-7) & 7.39(-7) \\ 
4  & 4& 1.13  & 1.71 & 0.1077 & 0.12105  & 8  & 0.88 & 0.775 & 3.17)-7) & 2.80(-7) \\ 
5  & 4& 1.71  & 2.88 & 0.1122 & 0.1248   & 4  & 0.94 & 0.775 & 8.87(-8) & 8.39(-8) \\ 
6  & 4& 2.88  & 4.10 & 0.1224 & 0.15215  & 4  & 0.95 & 0.775 & 2.73(-8) & 2.53(-8) \\ 
7  & 4& 4.10  & 5.90 & 0.1337 & 0.16875  & 5  & 0.97 & 0.775 & 1.60(-8) & 1.20(-8) \\ 
8  & 4& 5.90  & 11.9 & 0.1423 & 0.1719   & 4  & 1.00 & 1.0   & 3.17(-9) & 1.33(-9) \\ 
9  & 1& 11.9  & 20   & 0.1623 & 0.19925  & 1  & 1.00 & 1.0   & 3.66(-10)& 1.62(-10) \\ 
            \noalign{\smallskip} 
            \hline 
         \end{array} 
      \] 
\begin{list}{}{} 
\item[$N_{Cl}$ is the number of sample clusters per bin.]  
\item[$L_X(min)$, $L_X(max)$, $z_{min}$, and $z_{max}$ give the luminosity (in  
$10^{44}$ erg s$^{-1}$ [0.1 - 2.4 keV]) and redshift boundaries of the bins, respectively.]  
\item[$N_{tot}$ is the total number of clusters per bin including the unobserved ones.] 
\item[$^a)$ gives the mean volume coverage fraction of the REFLEX survey above the flux limit, for the detection of 30 photons for the bins.]  
\item[$^b)$ fraction of sky coverage for an interstellar column  
density $N_H \le 6\times 10^{20}$ cm$^{-2}$.] 
\item[$^{c)}$ cluster density (Mpc$^{-3}$) determined from the selection 
function derived in section 2.2, where the number in brackets gives 
the exponent of 10.]   
\item[$^{d)}$ cluster density (Mpc$^{-3}$) from the alternative method 
  used to determine the selection function.]  
\end{list} 
\end{table*}

\begin{table*} 
      \caption{The luminosity-redshift bins used for the alternatively 
        constructed test selection function.}  
         \label{tab5} 
      \[ 
         \begin{array}{lllllllll} 
            \hline 
            \noalign{\smallskip} 
 {\rm Bin~no.} & N_{Cl} & L_X (min) & L_X (max) & z_{min} & z_{max} & z (excl.) & z (excl.) & z (excl.) \\ 
(1) & (2) & (3) & (4) & (5) & (6) & (7) & (8) & (9) \\ 
            \noalign{\smallskip} 
            \hline 
            \noalign{\smallskip} 
1  & 4& 0.75 & 1.0 & 0.0550  & 0.06215 &  &  &  \\ 
2  & 4& 1.0  & 1.4 & 0.0794  & 0.08725 &  &  &  \\ 
3  & 4& 1.4  & 2.0 & 0.0920  & 0.1041  & (0.0926 - 0.09455) & (0.09665 - 0.09795) & (0.09865 - 0.09985) \\ 
4  & 4& 2.0  & 3.0 & 0.1077  & 0.12105 & (0.1102 - 0.1131) & (0.11835 - 0.1195) &  \\ 
5  & 4& 3.0  & 5.0 & 0.11145 & 0.1248  &  &  & \\ 
6  & 4& 5.0  & 7.0 & 0.1195  & 0.15215 & (0.1195 - 0.15685)^{*} &  & \\ 
7  & 4& 7.0  & 10  & 0.1161  & 0.16875 & (0.15135 - 0.16065) &  & \\ 
8  & 4& 10   & 20  & 0.0     & 0.1719  &  &  &  \\ 
9  & 1& 20   & 32  & 0.0     & 0.19925 &  &  &  \\   
            \noalign{\smallskip} 
            \hline 
         \end{array} 
      \] 
\begin{list}{}{} 
\item[The first 6 columns have the same meaning as those in Table 2, 
  where the luminosity bins are now given for a critical density] 
\item[Universe with a Hubble constant of $h_{100}=0.5$.] 
\item[$z_{excl}$ give the redshift intervals of the regions to be 
  excluded due to unobserved clusters scattered into the bins.] 
\item[$^{*)}$ gives the total redshift interval of the bins for a 
  Universe with critical density.]  
\end{list} 
\end{table*}

 
Two further selection criteria are important: (i) to avoid importing 
 galaxy clusters with lower quality detection 
parameters (flux error, extent parameter, etc.) we have only included 
galaxy clusters which contained more than 30 detected counts in the 
ROSAT All-Sky Survey. The same cut was made in the 
construction of the X-ray luminosity function in B\"ohringer et 
al. (2002). In addition, (ii) to obtain good X-ray spectra with a wide 
spectral coverage, we only selected clusters in sky areas where the 
hydrogen column density, $N_H$, measured at 21 cm (Dickey \& Lockman 
1990), is smaller than $6 \times 10^{20}$ cm$^{-2}$. This criterion was 
not applied to the most luminous clusters in bin number 8, since there 
are only few such objects. In addition the spectra of these clusters 
have high expected X-ray temperatures, and  will be less influenced by the 
hydrogen column density than those with lower temperatures. 
 
 
To determine the selection volume associated with each cluster we 
apply the following steps.  To take into account of the $N_H$ 
selection, we have inspected the fraction of the sky region in the 
REFLEX area with $N_H \ge 6 \times 10^{20}$ cm$^{-2}$.  This fraction 
is slightly dependent on the flux limit of the sky region, such that 
less sensitive regions have on average higher column densities. The 
fraction of sky area above our $N_H$-cut is about $22 - 23\%$ over 
more than 90\% of the sky. 
Only in a smaller less sensitive area is it slightly larger. Thus we 
correct the sky coverage by a factor of 0.775 for all bins except for 
bin 8 and 9, as shown in column 9 of Table 1.  
 
The other condition, 
that we should have a detection of at least 30 photons, further reduces 
the sky coverage, since in only 78\% of the REFLEX Survey area is the 
nominal flux limit reached for the detection of 30 photons (as 
explained in detail in the REFLEX sample construction paper, B\"ohringer et 
al. 2001). This effect is especially important for those bins which 
are close to the nominal flux limit. Therefore we have to determine 
the mean ``volume coverage'' of each bin as a function of the 
luminosity and redshift range within the bin.\footnote{Each bin 
constitutes an almost volume limited subsample. The ROSAT Survey 
contains, however, a few regions with reduced sensitivity, where the 
survey becomes flux-limited. Therefore each grid point of luminosity 
and redshift within the bin has a ``sky coverage'' of slightly less than 
100\%. The fractional sky coverage averaged over all $L_x$, $z$ grid 
points of the bin yields the mean ``volume coverage''.}  This average 
volume coverage per bin is different for each  
bin and the correction factors are given in Table 1. Both corrections 
are small but significant, so that it is important to include them.   
The empirical nature of these corrections will introduce only 
minor second order uncertainties, which are definitely only of the 
order of one percent.  
 
The final step is the normalization of the selection function. Most 
bins contain only the four clusters we initially selected. In this 
case we determine the cluster density for this luminosity bin by the 
inverse volume, multiplied by four (no multiplication in the case of 
bin 9).  In bins 3, 4, and 7, where we find new, unobserved 
clusters in the reconstructed sample due to redshift scattering, 
we normalize by the total number of clusters in the bin. 
 
The information provided for the selection function makes it possible 
to determine the distribution function of any property of the clusters 
in the sample. The estimated density of clusters for a given 
luminosity interval is the inverse of the selection volume multiplied 
by the number of clusters found in the luminosity bin. This 
calculation can also be restricted to a specific type of cluster. 
In this case for the density calculation the inverse volume is simply 
multiplied by the number of clusters of this type in the luminosity 
bin.  We will demonstrate how this is done for the luminosity function 
as an example in section 7. For further work with the {\sl 
REXCESS} sample, the construction of the temperature function, the 
mass function, but also more peculiar functions like the cool core gas 
mass function, this procedure will be important. This range of 
applications is precisely the strength of the XMM-Newton Legacy Program. 
The dataset provides, for the first time, a representative X-ray cluster 
sample, observed deep enough to provide a wealth of parameters on 
cluster structure (allowing for a complete cluster coverage by the 
XMM-Newton field-of-view), and at the same time large enough to allow 
the construction of meaningful statistics. 
 
\subsection{Alternative experimental sample selection} 
 
The proper statistical modeling of such a survey selection function is 
only to first order approximation a trivial task. In the presence of 
substantial measurement errors, or correlation uncertainties for a 
cluster property distribution function other than the luminosity 
function, scattering effects have to be accounted for. These effects 
correspond to the so-called Malmquist bias in flux limited samples. In 
the present case these effects are more complicated, and are best 
treated by Monte Carlo simulations. For a proper accounting one may 
not only consider the boundary migration effects, but also the 
boundary selection itself, since e.g. the midpoint rule selection 
depends on the statistics of the cluster distribution in luminosity 
and redshift space. This should also be included in the Monte Carlo 
simulations. 
 
Details of such an analysis will be considered in a future paper 
concerning the temperature or mass function construction. Here we 
adopt a didactical point of view, and attempt to illustrate the 
variance in the results due to the statistics of the boundary 
selection criterion, with an 
alternative selection scheme for the same clusters. 
 
To illustrate the robustness of the approach in the presence of sample 
variance effects in the sample selection, we adopt for test purposes a 
different variant of the above selection scheme. We cut the bins in 
redshift space around the observed clusters, now using the midpoint rule 
on both sides of the bin. The resulting alternative bin boundaries 
are listed in Table 2 (this Table also gives the luminosity values for 
the originally-chosen luminosity bins, defined for $H_0 = 50$ km s$^{-1}$ 
Mpc$^{-1}$ and a critical density universe).  Note that now bins 8 and 
9 begin at redshift $z=0$, since there are no lower redshift clusters 
with such a high X-ray luminosity in REFLEX. 
 
For the extra, unobserved clusters in the bins, we again apply again 
the midpoint rule to exclude the regions containing the unobserved 
clusters. The cut-out zones in redshift space for this recipe are given in 
Table 2. The cluster densities obtained with this second method are also   
given in Table 1. We discuss the effect of the two different ways of 
defining the sample selection function in Section 7, where we use these 
data to construct the luminosity function for this  sample.

\section{The sample} 
 
In total 34 galaxy clusters were selected from the REFLEX catalogue 
for this study, as listed in Table 3.  One of the selected objects, 
RXJ1350.7-3343, was found to have a purely point source X-ray emission 
from an AGN in the XMM-Newton images.  In the RASS its 
X-ray emission was found to be significantly extended (visual 
inspection confirmed by the KS test).  The origin of this extent is 
unclear, but could possibly be due to some attitude error in the data set that 
comes from different orbits. Such errors, if they occur at all in the 
RASS, must be very rare, since most of the known point like sources as 
stars and AGN do not feature such an extent. Therefore this object was 
removed from our sample and we have not included it in the above 
selection function construction. It is however listed in Table 3. 
 
Table 3 gives information on the X-ray properties of the 33 clusters 
and the AGN X-ray source as determined from the ROSAT All-Sky Survey 
data.  The columns of the table provide the following information: (1) 
the REFLEX name, (2) name given by Abell (1958) and Abell, Corwin \& 
Olowin (1989), (3) and (4) the right ascension and declination for the 
epoch J2000 in hours (degrees), minutes, and seconds, (5) the 
redshift, (6) the number of cluster galaxies from which the redshift 
has been determined, (7) and (8) the measured, unabsorbed X-ray flux, 
$F_x$, in units of $10^{-12}$ erg s$^{-1}$ cm$^{-2}$ for the 0.1 - 2.4 
keV energy band and the fractional error in percent\footnote{The 
fluxes and luminosities quoted here are those measured in the ROSAT 
All-Sky Survey.}, (9) the X-ray luminosity in units of $10^{44}$\,erg 
s$^{-1}$ in the rest frame 0.1 to 2.4  
keV band (uncorrected for missing 
flux), (10) the aperture radius in arcmin within which the X-ray 
count rate and flux were determined (the radius where the plateau 
value is reached in the cumulative count rate growth 
curve analysis), (11) the 0.1 - 2.4 keV luminosity corrected for the 
estimated flux lost outside the measurement aperture (by extrapolating 
to a radius of 12 core radii by means of a $\beta$-model with $\beta = 
2/3$, see B\"ohringer et al. 2004 for more details), (12) the 
interstellar HI column density in units of $10^{20}\,{\rm cm}^{-2}$ 
from Dickey \& Lockman (1990), and (13) the luminosity bin number to 
which the cluster belongs. 
 
To provide a complete documentation, we also list in Table 4 those 
clusters which were scattered into the sample bins in 
luminosity-redshift space due to the reconstruction of our 
sample. They are not observed in this project but are statistically 
accounted for. This table is similar in structure to Table 3 and the 
parameter description is the same. 
 
   \begin{figure} 
   \centering 
   \includegraphics[width=\columnwidth]{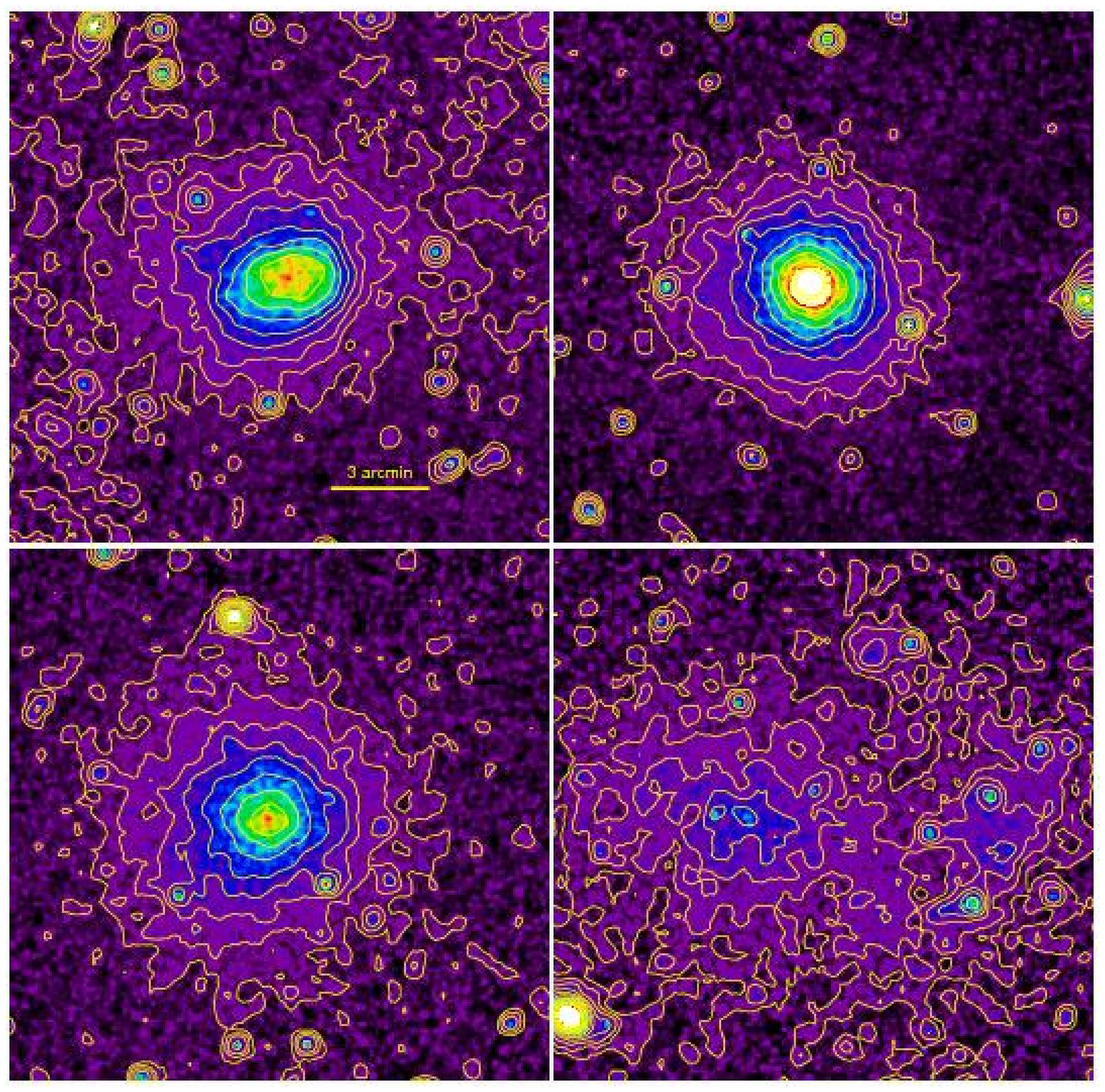} 
      \caption{Combined XMM-Newton MOS/pn 0.5 - 2 keV images of 
      the clusters in luminosity bin 1 ($4.07 - 5.5 \times 10^{43}$ 
      erg s$^{-1}$), RXCJ0345.7-4112 (S384, upper left), 
      RXCJ0225.1-2928 (upper right), RXCJ2023.0-2056 (S868, lower 
      left), and RXCJ2157.4-0747 (A2399, lower right). The images have 
      been corrected for vignetting and detector gaps, and the surface 
      brightness of the combined image has been normalized to that of 
      the pn detector. The background (not subtracted) is typically at 
      a level of $ 4 - 4.5 \times 10^{-3}$ cts s$^{-1}$ 
      arcmin$^{-2}$. The electronic version of the paper provides 
      colour versions of these images. The scale of the image is 
      marked by a 3 arcmin long bar. The contours start at a surface 
      brightness of $ 7.9 \times 10^{-3}$ cts s$^{-1}$ arcmin$^{-2}$ 
      and increase in steps of $\sqrt2$.} 
         \label{Fig2} 
   \end{figure} 
 
   \begin{figure} 
   \centering 
   \includegraphics[width=\columnwidth]{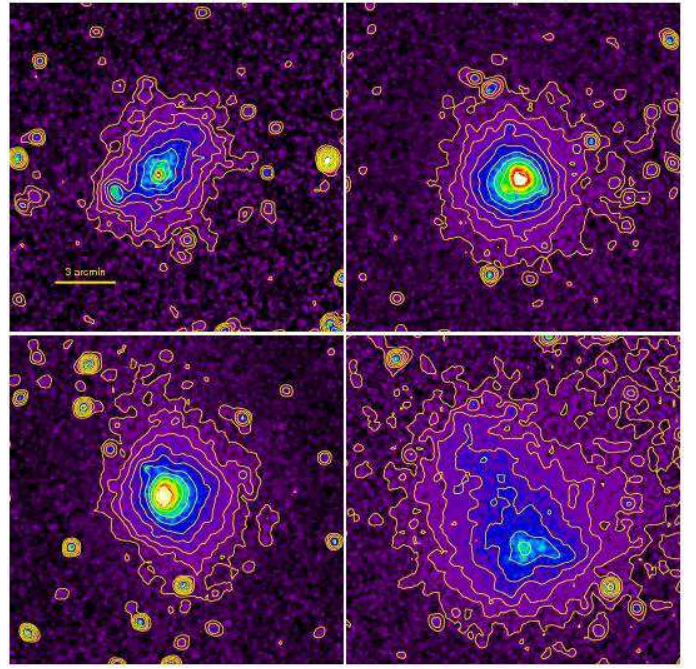} 
      \caption{Combined XMM-Newton MOS/pn 0.5 - 2 keV images of 
      the clusters in luminosity bin 2 ($5.5 - 7.8 \times 10^{43}$ erg 
      s$^{-1}$), RXCJ0821.8+0112 (A653, upper left), RXCJ1236.7-3354 
      (S700, upper right), RXCJ1302.8-0230 (A1663, lower left), and 
      RXCJ2129.8-5048 (A3771, lower right). The details are the same 
      as in Fig. 2.} 
         \label{Fig3} 
   \end{figure} 
 
   \begin{figure} 
   \centering 
   \includegraphics[width=\columnwidth]{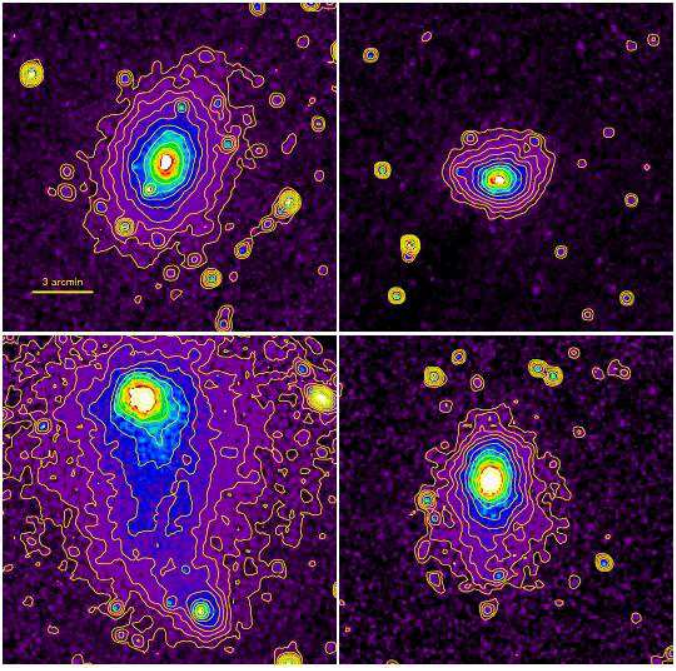} 
      \caption{Combined XMM-Newton MOS/pn 0.5 - 2 keV images of 
      the clusters in luminosity bin 3 ($0.78 - 1.13 \times 10^{44}$ erg 
      s$^{-1}$), RXCJ0003.8+0203 (A2700, upper left), RXCJ0211.4-4017 
      (A2984, upper right), RXCJ2152.2-1942 (A2384B, lower left), and 
      RXCJ2319.6-7313 (A3992, lower right). The details are the same 
      as in Fig. 2.} 
         \label{Fig4} 
   \end{figure} 
 
   \begin{figure} 
   \centering 
   \includegraphics[width=\columnwidth]{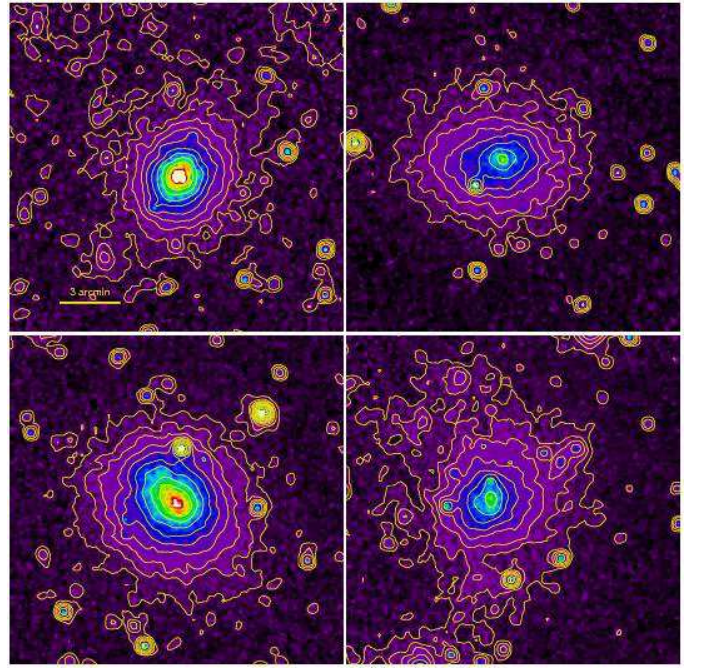} 
      \caption{Combined XMM-Newton MOS/pn 0.5 - 2 keV images of 
      the clusters in luminosity bin 4 ($1.13 - 1.71 \times 10^{44}$ erg 
      s$^{-1}$), RXCJ0049.4-2931 (S84, upper left), RXCJ0616.8-4748 
      (upper right), RXCJ1516.3+0005 (A2050, lower left), and 
      RXCJ1516.5-0056 (A2051, lower right). The details are the same 
      as in Fig. 2.} 
         \label{Fig5} 
   \end{figure} 
 
   \begin{figure} 
   \centering 
   \includegraphics[width=\columnwidth]{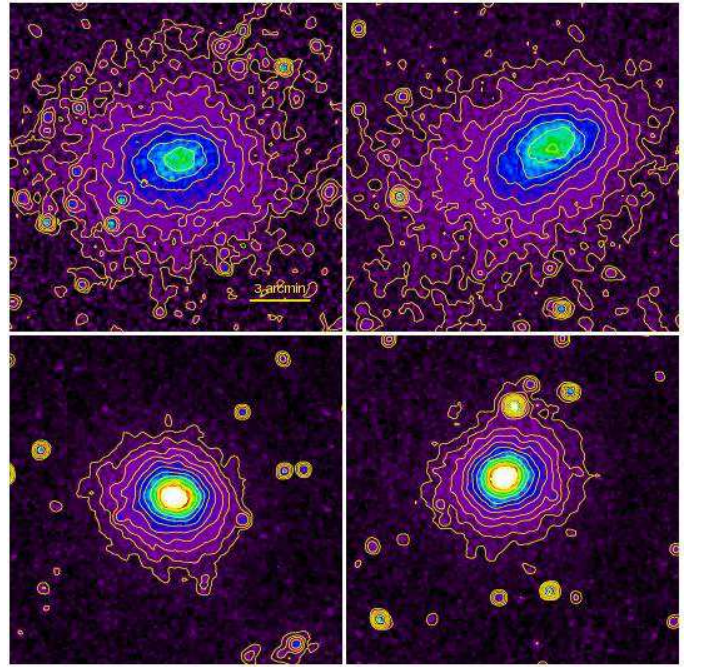} 
      \caption{Combined XMM-Newton MOS/pn 0.5 - 2 keV images of the 
      clusters in luminosity bin 5 ($1.71 - 2.88 \times 10^{44}$ erg 
      s$^{-1}$), RXCJ0006.0-3443 (A2721, upper left), RXCJ0145.0-5300 
      (A2941, upper right), RXCJ1141.4-1216 (A1348, lower left), and 
      RXCJ2149.1-3041 (A3814, lower right).The image of RXCJ0145.0-5300
      was produced from AO5 data, since none of the previous observations
      were clean enough to produce a decent image.} 
         \label{Fig6} 
   \end{figure} 
 
   \begin{figure} 
   \centering 
   \includegraphics[width=\columnwidth]{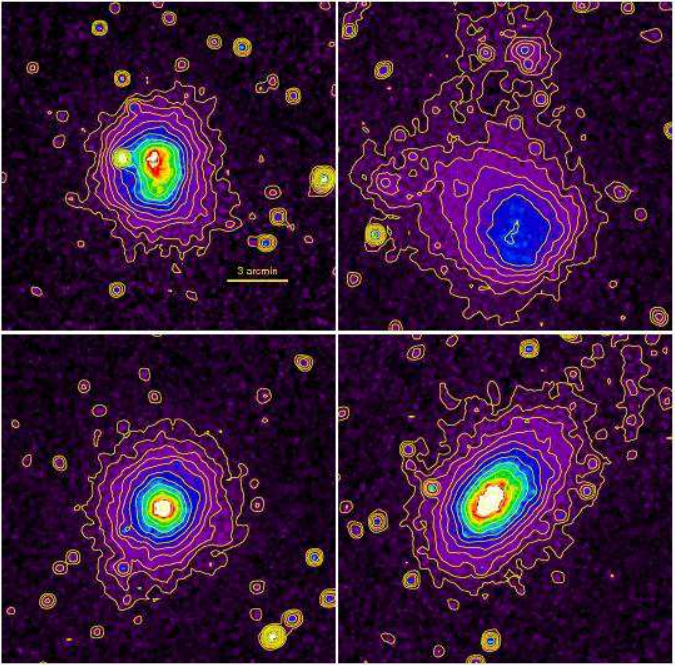} 
      \caption{Combined XMM-Newton MOS/pn 0.5 - 2 keV images of the 
      clusters in luminosity bin 6 ($2.88 - 4.10 \times 10^{44}$ erg 
      s$^{-1}$), RXCJ0020.7-2542 (A22, upper left), RXCJ2048.1-1750 
      (A2328, upper right), RXCJ2217.7-3543 (A3854, lower left), and 
      RXCJ2218.6-3853 (A3856, lower right). The details are the same 
      as in Fig. 2.} 
         \label{Fig7} 
   \end{figure} 
 
   \begin{figure} 
   \centering 
   \includegraphics[width=\columnwidth]{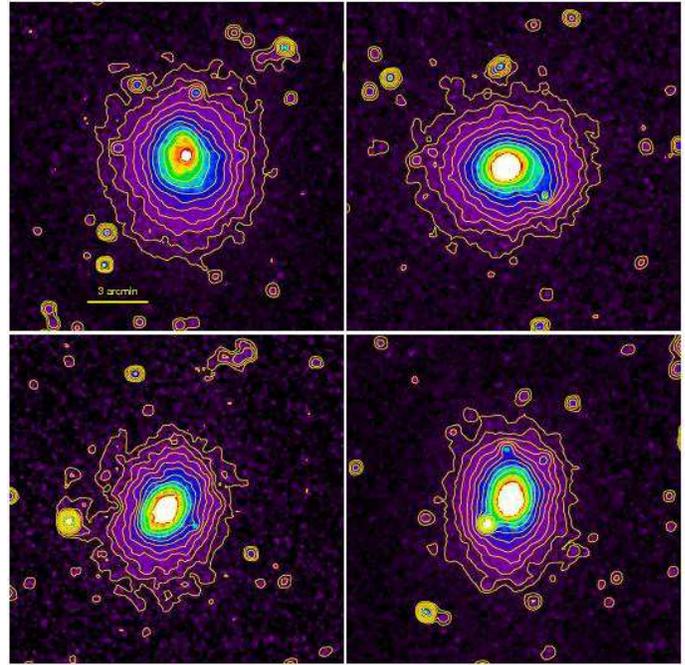} 
      \caption{Combined XMM-Newton MOS/pn 0.5 - 2 keV images of the 
      clusters in luminosity bin 7 ($4.10 - 5.90 \times 10^{44}$ erg 
      s$^{-1}$), RXCJ0547.6-3152 (A3364, upper left), RXCJ0605.8-3518 
      (A3378, upper right), RXCJ0958.3-1103 (A907, lower left), and 
      RXCJ1044.5-0704 (A1084, lower right). The details are the same 
      as in Fig. 2.} 
         \label{Fig8} 
   \end{figure} 
 
   \begin{figure} 
   \centering 
   \includegraphics[width=\columnwidth]{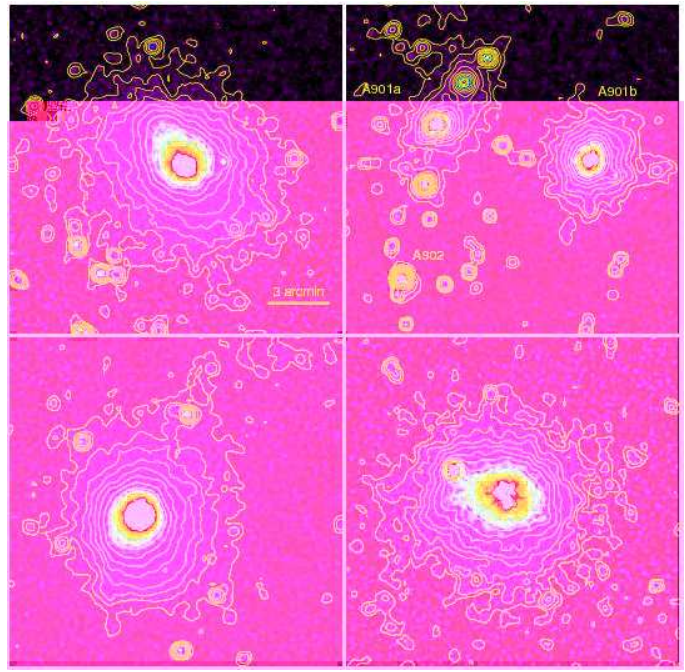} 
      \caption{Combined XMM-Newton MOS/pn 0.5 - 2 keV images of the 
      clusters in luminosity bin 8 ($5.90 - 11.9 \times 10^{44}$ erg 
      s$^{-1}$), RXCJ0645.4-5413 (A3404, upper left), RXCJ0956.4-1004 
      (A901, upper right), RXCJ2014.8-2430 (lower left), and 
      RXCJ2234.5-3744 (A3888, lower right). The details are the same 
      as in Fig. 2.  For the image of RXCJ0956.4-1004 only data from 
      the two MOS detectors have been used, since the pn 
      detector was closed during the observation.} 
         \label{Fig9} 
   \end{figure} 
 
   \begin{figure} 
   \centering 
   \includegraphics[width=\columnwidth]{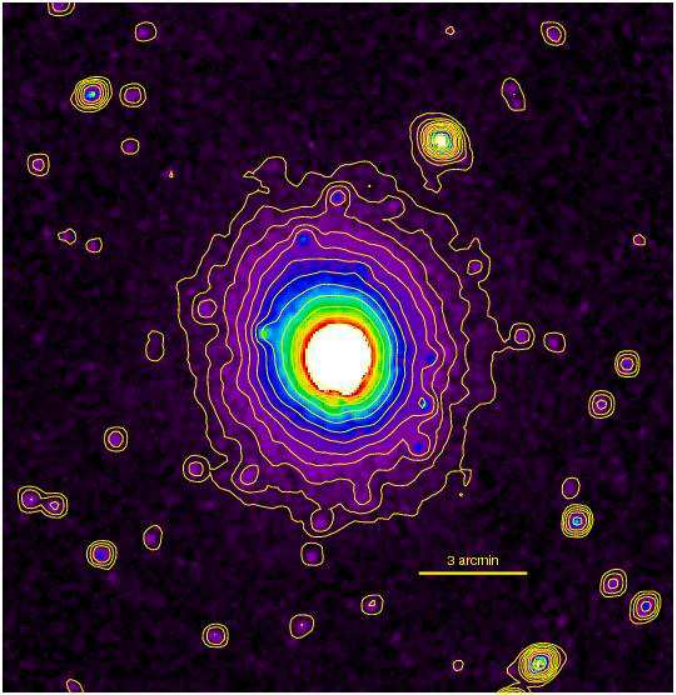} 
      \caption{Combined XMM-Newton MOS/pn 0.5 - 2 keV images of the 
      clusters in luminosity bin 9 ($11.9 - 20 \times 10^{44}$ erg s$^{-1}$). 
       The details are the same as in Fig. 2.} 
         \label{Fig10} 
   \end{figure} 

\begin{table*} 
      \caption{The REFLEX XMM Large Program cluster sample.} 
         \label{tab5} 
      \[  
         \begin{array}{llrrrrrrrrrrrl} 
            \hline 
            \noalign{\smallskip} 
 {\rm name}& {\rm alt. name}& RA(2000) & Dec(2000) & z & N_{gal}.& F_x & Error 
& L_x & R_{ap}& L_x^* & N_H & L_x-{\rm bin}  \\ 
(1) & (2) & (3) & (4) & (5) & (6) & (7) & (8) & (9) & (10) & (11) & (12) & (13) \\ 
            \noalign{\smallskip} 
            \hline 
            \noalign{\smallskip} 
 {\rm RXCJ0003.8+0203}&{\rm A2700       }&00~03~50.6& +02~03~48&0.0924&  9&  4.155&  18.8&  0.855&  8.5&  0.929&    3.0&  3  \\
 {\rm RXCJ0006.0-3443}&{\rm A2721       }&00~06~03.0& -34~43~27&0.1147& 75&  5.832&  13.6&  1.875& 10.0&  1.995&    1.2&  5  \\
 {\rm RXCJ0020.7-2542}&{\rm A0022       }&00~20~42.8& -25~42~37&0.1410&  3&  5.910&  12.1&  2.909&  7.5&  3.232&    2.3&  6  \\
 {\rm RXCJ0049.4-2931}&{\rm S0084       }&00~49~24.0& -29~31~28&0.1084& 18&  5.228&  16.0&  1.503& 11.0&  1.566&    1.8&  4  \\
 {\rm RXCJ0145.0-5300}&{\rm A2941       }&01~45~02.3& -53~00~50&0.1168&  4&  6.028&  16.0&  2.005&  7.5&  2.253&    2.3&  5  \\
 {\rm RXCJ0211.4-4017}&{\rm A2984       }&02~11~25.5& -40~17~12&0.1008&  6&  3.222&  11.5&  0.798&  8.0&  0.858&    1.4&  3  \\
 {\rm RXCJ0225.1-2928}&{\rm             }&02~25~10.5& -29~28~26&0.0604& 17&  4.736&  23.4&  0.408& 12.0&  0.434&    1.7&  1  \\
 {\rm RXCJ0345.7-4112}&{\rm S0384       }&03~45~45.7& -41~12~27&0.0603&  1&  5.763&  18.9&  0.495& 11.5&  0.532&    1.9&  1  \\
 {\rm RXCJ0547.6-3152}&{\rm A3364       }&05~47~38.2& -31~52~31&0.1483& 10&  8.526&   7.5&  4.667& 17.5&  4.667&    2.0&  7  \\
 {\rm RXCJ0605.8-3518}&{\rm A3378       }&06~05~52.8& -35~18~02&0.1392&  2&  9.393&   6.2&  4.478& 12.5&  4.665&    4.3&  7  \\
 {\rm RXCJ0616.8-4748}&{\rm             }&06~16~53.6& -47~48~18&0.1164&  1&  4.813&   9.8&  1.597& 14.0&  1.613&    4.8&  4  \\
 {\rm RXCJ0645.4-5413}&{\rm A3404       }&06~45~29.3& -54~13~08&0.1644&  2& 10.597&   7.9&  7.139& 13.0&  7.360&    6.6&  8  \\
 {\rm RXCJ0821.8+0112}&{\rm A0653       }&08~21~51.7& +01~12~42&0.0822&  6&  4.142&  19.2&  0.673& 12.0&  0.701&    4.2&  2  \\
 {\rm RXCJ0956.4-1004}&{\rm A0901/2     }&09~56~26.4& -10~04~12&0.1634&  9&  9.115&   9.6&  6.077& 17.0&  6.077&    5.1&  8  \\
 {\rm RXCJ0958.3-1103}&{\rm A0907       }&09~58~22.1& -11~03~35&0.1669&  2&  7.833&   8.3&  5.472&  8.5&  5.948&    5.1&  7  \\
 {\rm RXCJ1044.5-0704}&{\rm A1084       }&10~44~33.0& -07~04~22&0.1342&  6&  9.451&  12.2&  4.213&  7.0&  4.899&    3.4&  7  \\
 {\rm RXCJ1141.4-1216}&{\rm A1348       }&11~41~24.3& -12~16~20&0.1195&  6&  5.344&  12.3&  1.877&  7.0&  2.109&    3.3&  5  \\
 {\rm RXCJ1236.7-3354}&{\rm S0700       }&12~36~44.7& -33~54~10&0.0796&  4&  4.932&  19.8&  0.749&  8.5&  0.832&    5.6&  2  \\
 {\rm RXCJ1302.8-0230}&{\rm A1663       }&13~02~50.7& -02~30~22&0.0847&  3&  4.460&  21.9&  0.772& 12.0&  0.804&    1.7&  2  \\
 {\rm RXCJ1311.4-0120}&{\rm A1689       }&13~11~30.0& -01~20~07&0.1832& 66& 15.332&   8.0& 13.088& 10.5& 14.073&    1.8&  9  \\       
 {\rm RXJ1350.7-3343 }&{\rm AGN         }&13~50~43.9& -33~43~17&0.1142&  1&  4.021&  17.3&  1.280&  7.0&  1.422&    4.8&  4  \\
 {\rm RXCJ1516.3+0005}&{\rm A2050       }&15~16~19.2& +00~05~52&0.1181& 17&  4.956&  13.8&  1.697&  7.5&  1.886&    4.6&  4  \\
 {\rm RXCJ1516.5-0056}&{\rm A2051       }&15~16~34.0& -00~56~56&0.1198&  7&  3.918&  61.7&  1.383& 13.0&  1.397&    5.5&  4  \\
 {\rm RXCJ2014.8-2430}&{\rm             }&20~14~49.7& -24~30~30&0.1612&  2& 14.040&  13.4&  9.157&  6.5& 11.033&    7.4&  8  \\
 {\rm RXCJ2023.0-2056}&{\rm S0868       }&20~23~01.6& -20~56~55&0.0564&  2&  5.497&  17.8&  0.411&  8.5&  0.467&    5.6&  1  \\
 {\rm RXCJ2048.1-1750}&{\rm A2328       }&20~48~10.6& -17~50~38&0.1475&  3&  5.930&  15.2&  3.215& 10.5&  3.349&    4.8&  6  \\
 {\rm RXCJ2129.8-5048}&{\rm A3771       }&21~29~51.0& -50~48~04&0.0796&  2&  5.051&  66.2&  0.767& 11.5&  0.807&    2.2&  2  \\
 {\rm RXCJ2149.1-3041}&{\rm A3814       }&21~49~07.4& -30~41~55&0.1184& 19&  6.182&  12.9&  2.117&  9.5&  2.276&    2.3&  5  \\
 {\rm RXCJ2152.2-1942}&{\rm A2384 (B)   }&21~52~14.2& -19~42~20&0.0963&  4&  4.059&  30.0&  0.912&  6.0&  1.060&    3.0&  3  \\
 {\rm RXCJ2157.4-0747}&{\rm A2399       }&21~57~25.8& -07~47~41&0.0579&  8&  5.851&  19.0&  0.462& 14.0&  0.481&    3.5&  1  \\
 {\rm RXCJ2217.7-3543}&{\rm A3854       }&22~17~43.3& -35~43~34&0.1486& 44&  6.406&  10.7&  3.535&  8.5&  3.842&    1.1&  6  \\
 {\rm RXCJ2218.6-3853}&{\rm A3856       }&22~18~40.2& -38~53~51&0.1411& 10&  7.132&  10.2&  3.516&  9.0&  3.781&    1.3&  6  \\
 {\rm RXCJ2234.5-3744}&{\rm A3888       }&22~34~31.0& -37~44~06&0.1510& 70& 11.225&   8.9&  6.363&  7.5&  7.314&    1.2&  8  \\
 {\rm RXCJ2319.6-7313}&{\rm A3992       }&23~19~41.8& -73~13~51&0.0984&  3&  3.993&  17.9&  0.937&  7.5&  1.030&    1.9&  3  \\
 
            \noalign{\smallskip} 
            \hline 
         \end{array} 
      \] 
\end{table*} 
\begin{table*} 
      \caption{The additional clusters contained in the redshift-luminosity bins, which are 
        not part of the observed cluster sample with more than 30 detected X-ray counts in the RASS  
        and $N_H \le 6\times 10^{20}$ cm$^{-2}$.} 
         \label{tab5} 
      \[ 
         \begin{array}{llrrrrrrrrrrr} 
            \hline 
            \noalign{\smallskip} 
 {\rm name}& {\rm alt. name}& RA(2000) & Dec(2000) & z & N_{gal}.& F_x & Error 
& L_x & R_{ap}& L_x^* & N_H &  L_x-{\rm bin}  \\ 
(1) & (2) & (3) & (4) & (5) & (6) & (7) & (8) & (9) & (10) & (11) & (12) & (13) \\ 
            \noalign{\smallskip} 
            \hline 
            \noalign{\smallskip} 
 {\rm RXCJ0028.6-2338}&{\rm A0042       }&00~28~39.3& -23~38~14&0.1120&  5&  4.836&  13.3&  1.491& 13.0&  1.521&    1.8&   4 \\
 {\rm RXCJ0107.8-3643}&{\rm A2871       }&01~07~49.1& -36~43~38&0.1186& 19&  3.557&  13.6&  1.229& 12.0&  1.254&    1.9&   4 \\ 
 {\rm RXCJ0108.9-1537}&{\rm A0151S      }&01~08~55.2& -15~37~44&0.0970& 13&  3.703&  15.3&  0.845& 10.0&  0.889&    1.8&   3 \\   
 {\rm RXCJ0548.8-2154}&{\rm             }&05~48~50.4& -21~54~43&0.0928&  9&  3.977&  12.3&  0.825& 11.5&  0.859&    3.0&   3 \\
 {\rm RXCJ1038.4-2454}&{\rm             }&10~38~24.1& -24~54~10&0.1230& 10&  4.134&  13.5&  1.545&  8.5&  1.661&    5.5&   4 \\
 {\rm RXCJ1512.8-0128}&{\rm             }&15~12~51.0& -01~28~47&0.1223&  2&  3.354&  18.5&  1.238&  5.5&  1.423&    5.2&   4 \\
 {\rm RXCJ2220.5-3509}&{\rm A3866       }&22~20~34.6& -35~09~53&0.1544&  1&  9.489&   8.8&  5.656&  8.5&  6.215&    1.1&   7 \\
 {\rm RXCJ2359.3-6042}&{\rm A4067       }&23~59~19.2& -60~42~00&0.0989& 30&  4.544&  20.5&  1.080&  9.5&  1.149&    2.4&   3 \\
 
            \noalign{\smallskip} 
            \hline 
         \end{array} 
      \] 
\end{table*} 
 
Three of the clusters had previous XMM-Newton observations in the 
archive.  For A1689 (RXCJ1311.4-0120) and the A901/A902 cluster 
complex (RXCJ0956.4-1004), the exposure times were sufficient or 
longer than required for this study, and we thus could make use of the archive 
data. A3888 had only a very short exposure in the archive. Thus we 
complemented this observation by additional exposure time to bring the 
data to the same depth as for the other clusters. In our AO3 
proposal we successfully requested the observation of 32 
targets. Since about 37\% of the observations suffered from severe 
contamination by solar flares for a substantial part of the observing 
time, we requested the reobservation of 3 clusters in AO4 and 9 
further clusters in AO5. In this paper all the AO4 results are 
included. At the time of writing a large part, but not all, of the AO5 
observations were completed and their data analysis is ongoing. 
 
Table 5 provides an overview on the observation parameters and the 
data quality.  In addition to the observation numbers, dates and 
nominal observation times, we give the mean LIVETIMES of the 
detectors chips.  We only list the values for pn and MOS1, since the 
equivalent data for MOS2 are always very similar to those of MOS1.  In 
the final columns we list the exposure times left for scientific 
analysis after a cleaning of the data for soft proton flares. 
 
The cleaning done here, which is used to obtain a first overview on 
the data quality and to produce the image results shown below, is 
similar to that used in Pratt et al. (2007), where the data screening 
is optimized for the spectroscopic analysis. For a more detailed 
description of the screening we refer to that paper. In brief, we 
conduct a first data cleaning by means of a 3$\sigma$ clipping above 
the ``quiet level'' in the hard band light curves (12 - 14 keV for pn 
and 10 - 12 keV for MOS) in 100 s intervals, where the quiet level is 
characterized by a Gaussian distribution of the count rate at low 
count rate levels. In some cases where the observation is so disturbed 
that the Gaussian distribution is not easily established, we have used 
standard cut values, as noted in Table 5. The second screening is 
performed in a wider band (0.5 - 10 keV) in 10 s 
intervals. Most of the periods with high background are removed in the 
first cleaning step. The second stage mostly affects the flanks of the 
flares, and occasionally a flare which is very soft. Typically about 5 - 
10\% of the remaining data in flare-affected observations are removed 
in the second step.  The cut values listed in Table 5 refer to the 
second wide band cleaning, and are given in units of counts in the 
total detector in 10 s intervals.  We also remark on the effectiveness 
of this particular cleaning process for flagging data sets that are 
good, that have been cleaned by setting the cuts manually to a 
standard value, data which have an enhanced residual background, and 
data where one of the detectors is left with essentially no data. 
 
\section{X-ray images of the sample clusters} 
 
Fig. 2 to 10 show images of the clusters in the 0.5 - 2 keV band, an 
energy range which has an optimal signal-to-noise. The images are 
grouped by in bins of increasing luminosity. The images are 
produced from the cleaned event files, normalized by the exposure maps 
which include the vignetting correction, gaps and bad pixel 
information. The images of all detectors are combined with the pixel 
count rates of the two MOS detectors scaled to the pn sensitivity for 
a typical cluster spectrum. The combination is performed for the 
exposure maps and for the images separately, such that almost all gaps 
and bad pixel holes are filled by the information from at least one 
detector. The images are then smoothed by a Gaussian filter with a 
$\sigma$-width of 4 arcsec (which is slightly less than the instrument 
PSF). 
 
The colour scale of the images is scaled with a 
factor of $L_X^{0.22}$, for the following reason. In the simple 
self similar picture of clusters (see e.g. Kaiser 1986) we expect the 
central intracluster plasma density to be roughly constant (ignoring 
the known deviations due to secular entropy modifications), and 
also the gas density profile as a function of the scaled radius, 
$r/r_{500}$ should be roughly the same (e.g. Arnaud et al. 2002). 
Since the surface brightness is proportional to the plasma density 
squared integrated along the line of sight through the cluster, the 
surface brightness then scales only with the line-of-sight extent of 
the cluster, that is with $r_ {500}$. Taking $r_{500} \propto M^{1/3}$ 
and $L_X \propto M^{1.5}$ we obtain the above relation between X-ray 
luminosity and surface brightness. The scaling does not take into 
account the surface brightness dimming with redshift, however, 
although the redshift interval covered by these clusters is relatively 
small.  The surface brightness contours used in the figures were not 
scaled but start at a fixed ratio to the typical background and 
increase in logarithmic steps (by a factor of $\sqrt2$). 
 
With this scaling we readily recognize clusters with very dense cores 
(cooling cores) as those with very bright centres. Clusters which 
barely reach green colours (displayed in the electronic version of the 
paper) feature a very low surface brightness, indicating that these 
clusters are most probably dynamically young.

\begin{table*} 
      \caption{Overview of the XMM-Newton observation parameters of 
        the cluster sample, up to and including AO4.} 
         \label{tab56} 
      \[ 
         \begin{array}{llrrrrrrrrrr} 
            \hline 
            \noalign{\smallskip} 
 {\rm name} &{\rm observation} &{\rm date}  &{\rm nominal} &{\rm nominal}  &{\rm total}  &{\rm total}    &{\rm cleaned}&{\rm cleaned} 
   & {\rm cut} & {\rm cut} & {\rm flag} \\ 
            &{\rm number}      &{\rm (d.m.y)}&{\rm exp.~PN} &{\rm exp.~MOS1}&{\rm exp.~PN}&{\rm exp.~MOS1}&{\rm exp.~PN}&{\rm exp.~MOS1} 
   & {\rm PN}  & {\rm MOS1 } & \\ 
(1) & (2) & (3) & (4) & (5) & (6) & (7) & (8) & (9) & (10) & (11) & (12) \\ 
            \noalign{\smallskip} 
            \hline 
            \noalign{\smallskip} 
 {\rm RXCJ0003+0203 }& 201900101&24.06.04& 23279& 26667& 20242& 26223& 19409& 26002& 102&  37& 1 \\
 {\rm RXCJ0006-3443a}& 201900201&08.12.04& 17019& 15454&  8508&  3109&  1303&     0&  80&  33& 2 \\
 {\rm RXCJ0006-3443b}& 201903801&13.05.05& 13836& 17763& 11872& 17493&  5852& 12201& 113&  45& 3\\
 {\rm RXCJ0020-2542 }& 201900301&26.05.04& 26379& 29767& 23001& 29401& 10732& 15358&  93&  39& 1\\
 {\rm RXCJ0049-2931 }& 201900401&04.12.04& 31333& 34519& 22992& 33630& 13304& 19814& 159&  43& 3\\
 {\rm RXCJ0145-5300 }& 201900501&12.11.04& 25764& 27498& 17997& 27126&     0&   702&  80&  33& 2\\
 {\rm RXCJ0211-4017 }& 201900601&27.12.04& 25240& 29167& 21829& 28740& 21741& 28734&  83&  37& 1\\
 {\rm RXCJ0225-2928a}& 201900701&06.07.04& 25301& 29227& 18205&     0&  4234&     0&  82&   0& 0\\
 {\rm RXCJ0225-2928b}& 302610601&27.01.06& 22440& 26367& 19197& 26025& 16519& 20290& 198&  43& 3\\
 {\rm RXCJ0345-4112 }& 201900801&05.03.04& 23279& 26667& 20541& 26233&  8154& 17465&  94&  44& 1\\
 {\rm RXCJ0547-3152 }& 201900901&07.03.04& 21679& 25067& 19144& 24819& 17604& 23464&  97&  39& 1\\
 {\rm RXCJ0605-3518 }& 201901001&29.10.04& 20940& 26667& 18136& 26327& 14756& 20124& 116&  53&~1\\
 {\rm RXCJ0616-4748a}& 201901101&26.04.04& 25379& 27183& 17523& 26103&  3689&  6817&  78& 109& 3\\
 {\rm RXCJ0616-4748b}& 302610401&05.01.06& 23940& 27867& 20493& 27539& 18719& 22700&  81&  31& 1\\
 {\rm RXCJ0645-5413a}& 201901201&07.05.04& 19874& 18517&     0& 18328&     0& 11167&   0&  40& 0\\
 {\rm RXCJ0645-5413b}& 201903401&12.06.04& 17279& 20667& 14936& 20448&  4304&  6908&  80&  33& 2\\
 {\rm RXCJ0821+0112a}& 201901301&13.10.04& 22001& 15668& 16285&  7315&   173&  5921&  80&  33& 2\\
 {\rm RXCJ0821+0112b}& 201903601&15.11.04&  7740& 11667&  6775& 11545&  6765& 11469&  73&  30& 1\\
 {\rm RXCJ0956-1004 }& 148170101&06.05.03& 94321& 94333&  3438& 60448&  3438& 43202&  52&  27& 1\\
 {\rm RXCJ0958-1103a}& 201901401&09.05.04& 12031& 14950& 10251&  2038&  1933&  5000&  76&  50& 3\\
 {\rm RXCJ0958-1103b}& 201903501&17.06.04& 11279& 14667&  9763& 14433&  4830&  8588&  99&  43& 1\\
 {\rm RXCJ1044-0704 }& 201901501&23.12.04& 25240& 29167& 21833& 28827& 18293& 25712& 109&  45& 1\\
 {\rm RXCJ1141-1216 }& 201901601&09.07.04& 32805& 32274& 24343& 31653& 21920& 28274&  76&  30& 1\\
 {\rm RXCJ1236-3354a}& 201901701&28.07.04& 19274& 24268& 15693& 24026&   260&  6907&  80&  33& 2\\
 {\rm RXCJ1236-3354b}& 201903701&30.12.04& 12140& 16067& 10562& 15871&  9448& 13781& 102&  39& 1\\
 {\rm RXCJ1236-3354c}& 302610701&20.01.06& 20940& 24867& 17974& 24573& 17870& 24173&  83&  32& 1\\
 {\rm RXCJ1302-0230 }& 201901801&22.06.04& 20479& 25667& 17844& 25243& 16443& 24538&  84&  35& 1\\
 {\rm RXCJ1311-0120 }& 093030101&24.12.01& 34798& 39167& 30682& 38403& 29224& 36588& 134&  52&~1\\
 {\rm RXCJ1350-3343 }& 201901901&15.02.04& 22879& 26267& 20171& 25962&  5294& 14532&  80&  33& 2\\
 {\rm RXCJ1516+0005 }& 201902001&22.07.04& 24240& 28167& 21209& 27858& 21058& 26500& 105&  43& 1\\
 {\rm RXCJ1516-0056 }& 201902101&03.08.04& 26240& 30167& 23030& 29748& 21750& 29284&  83&  33& 1\\
 {\rm RXCJ2014-2430 }& 201902201&08.10.04& 22740& 26667& 19531& 26170& 16041& 24677& 137&  59& 3\\
 {\rm RXCJ2023-2056 }& 201902301&06.04.05& 25740& 29667& 21070& 29346&  9205& 17380&  93&  34& 1\\
 {\rm RXCJ2048-1750 }& 201902401&13.05.04& 23279& 26667& 20166& 26116& 18728& 25119&  85&  36& 1\\
 {\rm RXCJ2129-5048 }& 201902501&16.10.04& 21740& 25667& 18827& 25376& 12508& 23163& 123&  46& 3\\
 {\rm RXCJ2149-3041 }& 201902601&29.11.04& 22740& 26667& 19698& 26327& 18019& 25279&  69&  39& 1\\
 {\rm RXCJ2152-1942 }& 201902701&28.10.04& 22740& 26667& 19674& 26363& 10962& 21226& 111&  49& 1\\
 {\rm RXCJ2157-0747 }& 201902801&11.05.05& 22740& 24334& 18491& 17351&  7387& 10460&  98&  36& 1\\
 {\rm RXCJ2217-3543 }& 201902901&12.05.05& 22741& 26668& 19899& 26392& 16767& 23637&  90&  34& 1\\
 {\rm RXCJ2218-3853 }& 201903001&24.10.04& 24340& 28267& 20918& 27801& 12328& 22396& 129&  51&~1\\
 {\rm RXCJ2234-3744a}& 201903101&10.11.04& 26540& 30467& 11728& 30057&  4808&    59&  80&  33& 2\\
 {\rm RXCJ2234-3744b}& 018741701&03.05.01&  4488&  7114&  4070&  7050&  3952&  6836& 135&  53& 1\\
 {\rm RXCJ2319-7313a}& 201903201&18.04.04& 24827& 15681&  8980& 14987&     0&   279& 999& 565& 3\\
 {\rm RXCJ2319-7313b}& 201903301&15.05.04&  7279& 10667&  6368& 10543&  6363& 10277&  83&  35& 1\\
 
            \noalign{\smallskip} 
            \hline 
         \end{array} 
      \] 
The nominal exposure times (columns 4 \& 5) are obtained from the 
observation log browser of the XMM-Newton archive 
(http://xmm.vilspa.esa.es/external/xmm$_{-}$obs$_{-}$info/obs$_{-}$view$_{-}$frame.shtml) 
for each detecter.  The total exposure times (columns 6 \& 7) are the 
mean chip LIVETIMES read from the event file headers, and the cleaned 
times (columns 8 \& 9) are obtained after the application of the 
two-step cleaning process described in the text. Columns 10 \& 11 
give the cut values for the second, wider band cleaning in units of 
cts per 10s. The flag indicates good cleaning (1), cleaning with 
standard cuts (2), imperfect cleaning with residual high background 
sufficient for the image analysis but not necessarily for spectroscopy 
(3), and cases where the exposure for one of the detectors has 
effectively been lost (0). 
\end{table*}


\section{Remarks on some clusters} 
 
There will be at least one dedicated publication addressing in detail 
the morphology of the clusters in this sample. Here we briefly 
comment on some of the peculiar clusters. There are 2 clusters with  
multiple components, 3 clusters with a complex, diffuse,  
low surface brightness appearance, and one cluster where the data are 
still sparse. 
 
{\bf RXCJ2152.2-1942} (Fig. 4) was selected as the fainter southern 
component of this bimodal cluster. In the ROSAT All-Sky Survey the two 
emission regions appear almost distinct and the system was therefore 
split into two clusters (two separate dark matter halos) in the REFLEX 
catalogue. In the survey selection only the southern component should be 
counted. The deeper XMM-Newton exposure now reveals that the two 
systems are interacting. The total system is catalogued in the optical 
as A2384 by Abell (1958). 
 
{\bf RXCJ0956.4-1004} (Fig. 9), also known as A901a, A901b, and A902, 
is a system of several diffuse and point-like X-ray sources. In the 
ROSAT All-Sky Survey we observed a complex emission region that was 
catalogued as one object.  For our analysis we have used the archival 
XMM-Newton observation. The nominal observing time of this observation 
was very long, $\sim 94$ ksec, but only half of the MOS 
observing time is useful due to a series of strong flares, and the 
pn detector was closed during the observation. 
  
Gray et al. (2002) find three major mass concentrations, A901a, A901b, 
and A902, in their lensing analysis, and call the structure a 
supercluster at redshift $z = 0.16$.  Only A901b shows the  
extended, but compact, X-ray emission expected from a well evolved rich 
X-ray luminous cluster, as noted previously from the ROSAT HRI observation 
by Schindler (2000). The X-ray emission from A901a is dominated by a 
very strong point source, associated with a faint galaxy. There is 
definitely also extended emission associated with this mass 
component. The extended X-ray emission is centered on the central 
dominant elliptical of A901a in the west of the X-ray point source and 
very diffuse low surface brightness emission is observed on larger 
scale. The third mass component A902 is also associated with very 
diffuse low surface brightness emission, which indicates a dynamically 
very young galaxy cluster structure. The extended X-ray emission 
around A901a and A902 was not noted in the ROSAT HRI study by 
Schindler (2000), which involved much fewer photons. More details on 
the morphology of this cluster will be described in a forthcoming 
paper from our collaboration. 
 
{\bf RXCJ2157.4-0747} (Fig. 2), A2399, is a bimodal system with very 
diffuse, low surface brightness X-ray emission. Like the following two 
clusters, this is most probably a dynamically young object in 
formation without a significant preexisting cluster core. 
 
{\bf RXCJ 2129.8-5048} (Fig. 3), A3771, is another low surface 
brightness cluster, which is dynamically young, but does not feature a 
multi-component configuration. 
 
{\bf RXCJ2048.1-1750} (Fig. 7), A2328, is similar in its morphology to 
the previous cluster, but is more luminous and thus massive. In addition it 
features two smaller possibly infalling systems at its outskirts. 
 
{\bf RXCJ0145.0-5300} (Fig. 6), A2941, has insufficient data even for 
the production of a decent X-ray image. After flare cleaning there is 
no useful pn time and only a few hundred seconds of MOS 
exposure are left, from which the image has been made. The 
observation of this cluster is rescheduled in AO5. 
 
\section{Comparison between ROSAT All-Sky Survey and XMM-Newton fluxes} 
 
   \begin{figure} 
   \centering 
   \includegraphics[width=\columnwidth]{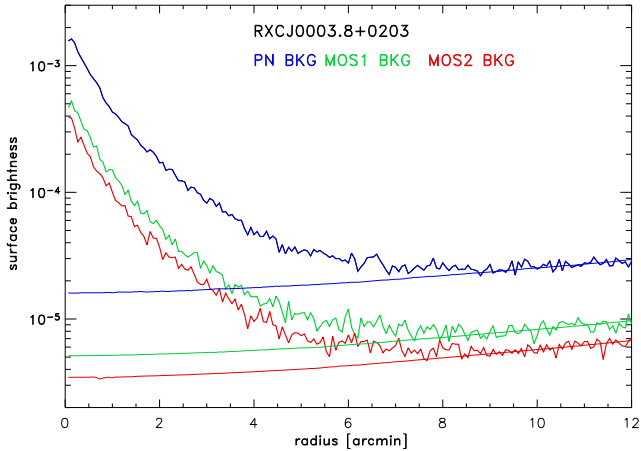} 
      \caption{[0.5 - 2 keV] surface brightness profile of the cluster 
      RXCJ0003.8+0203, for all three detectors, in units of counts per 
      $4\times4$ arcsec$^2$ pixel s$^{-1}$ (upper curves), plotted 
      with the scaled, modelled background surface brightness (lower 
      curves). The MOS2 surface brightness profile has been multiplied 
      by a factor of 0.7 for better visibility.  The background 
      surface brightness is increasing with radius because the 
      background is vignetting corrected, which overweights the 
      particle background in the outer regions. No significant cluster 
      emission is seen for this target at radii outside about 9 
      arcmin.} 
         \label{Fig11} 
   \end{figure} 
   \begin{figure} 
   \centering 
   \includegraphics[width=\columnwidth]{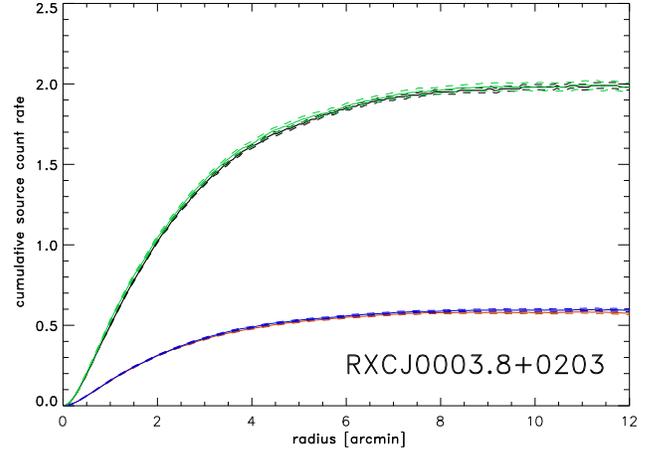} 
      \caption{Cumulative, background subtracted [0.5 - 2 keV] count 
      rates (``growth curves'') of the cluster RXCJ0003.8+0203 as a 
      function of the cluster radius.  The black upper curve (with 
      error corridors) refers to the pn results, and the two 
      lower curves with their errors corridors correspond to the two MOS 
      results, respectively.  The green, upper curve is the MOS1 
      growth curve scaled to the sensitivity of the pn. The curves 
      reach a flat plateau outside a radius of about 9 arcmin.} 
         \label{Fig12} 
   \end{figure} 
 
Using the present XMM-Newton observations we measured the fluxes from 
the galaxy clusters in the 0.5 - 2 keV band, and compared the results 
to the previous ROSAT All-Sky Survey (RASS) observations.  In the RASS 
the count rates from which the fluxes are derived were determined by 
the growth curve analysis technique described in B\"ohringer et 
al. (2000). Here we use a similar approach. 
 
We first construct cluster images for each detector. We then integrate 
the counts in concentric rings, weighting each image pixel by the 
vignetting corrected exposure maps. We excise all bad pixels and 
pixels which fall into gaps or low exposure regions near gaps, and 
correct for the area lost in the ring. To estimate the background 
contribution we use the background data provided by Read and Ponman 
(Read \& Ponman 2003) with the same cleaning as applied by Pratt et 
al. (2007), recast onto the same sky position and orientation as the 
target fields. In these data sets X-ray sources have been removed and 
the images produced from the data sets feature depressions in these 
removal zones.  We therefore apply a model fit to the background by 
means of the {\sl SAS} task {\sl esplinemap} with the parameter {\sl 
fitmethod = model}.  We compare the model background surface 
brightness distribution to the target data in the same outer region 
(where we have insignificant cluster contribution to the X-ray image) 
and scale the background to the image surface brightness in this 
region where the profiles have the same shape.  This scaled background 
is then subtracted from the cluster profile. An example of a cluster 
profile and the scaled background is shown in Fig. 11. We have tested 
the validity of this procedure by checking the change of the results 
as a function of the radius limit outside which the data are used for 
the renormalization and find very little change ($\le 1\%$) for 
limiting radii $\ge$ 9 or 10 arcmin, depending on the shape of the 
cluster. 
 
To account for the point source contribution to a cluster's X-ray 
emission we have also produced ''cleaned'' images in the following 
way. We have run the {\sl SAS} source detection procedure {\sl 
ewavelet} to localize point sources. Since also the cluster centers 
and substructure are usually recognized as X-ray sources by {\sl 
ewavelet} we have removed the detected sources through visual 
inspection, retaining all diffuse cluster emission including 
substructure and central cusps.  We used the radii of the {\sl 
ewavelet} algorithm in SAS as excision radii, in a first attempt to 
exclude the point sources. This radius is increased after visual 
inspection for the brighter sources which are not completely 
removed. The same regions are excised in the exposure maps. We use 
these and the uncleaned images to measure the cluster and the total 
flux in the cluster region, respectively. 
 
An integration of the surface brightness profiles times the area of 
the rings gives the count rate growth curves as shown in 
Fig. 12. These level off at large radii.  Fluxes are derived from 
these count rates by means of count rate to flux conversion factors 
determined using XSPEC 
software\footnote{http://heasarc.gsfc.nasa.gov/docs/xanadu/xspec}.  To 
determine these conversion factors a spectral model has to be defined 
in XSPEC. For 15 of the clusters, we use an absorbed MEKAL 
model\footnote{http://heasarc.gsfc.nasa.gov/docs/xanadu/xspec/ \\ 
manual/XSmodelMekal.html} with a temperature as measured by a single 
temperature fit to the data in the radial region 0.1 to 0.4 $r_{200}$ 
according to the analysis performed by Pratt et al. (2007). Accurate 
temperature measurements of this kind are not yet available for some 
clusters. For these remaining clusters we use a temperature estimate 
from the $L_X$-temperature relation defined by Eq. 1, with the 
interstellar column densities listed in Table 3 and metallicities of 
0.3 solar. The uncertainty in the flux conversion factor, even in the 
case of a factor 2 difference in the temperature estimate, is never 
larger than 3\%, and in most cases is much less. The error in the 
measured flux accounts for the Poisson error of the source counts as 
well as the photon statistical error in the background 
subtraction. More precise values for the fluxes and luminosities will 
be reported in a later paper when all the data are at hand and have 
been reduced. For these final results we will also consider the 
additional correction for the temperature variation with radius. 
 
   \begin{figure} 
   \centering 
   \includegraphics[width=\columnwidth]{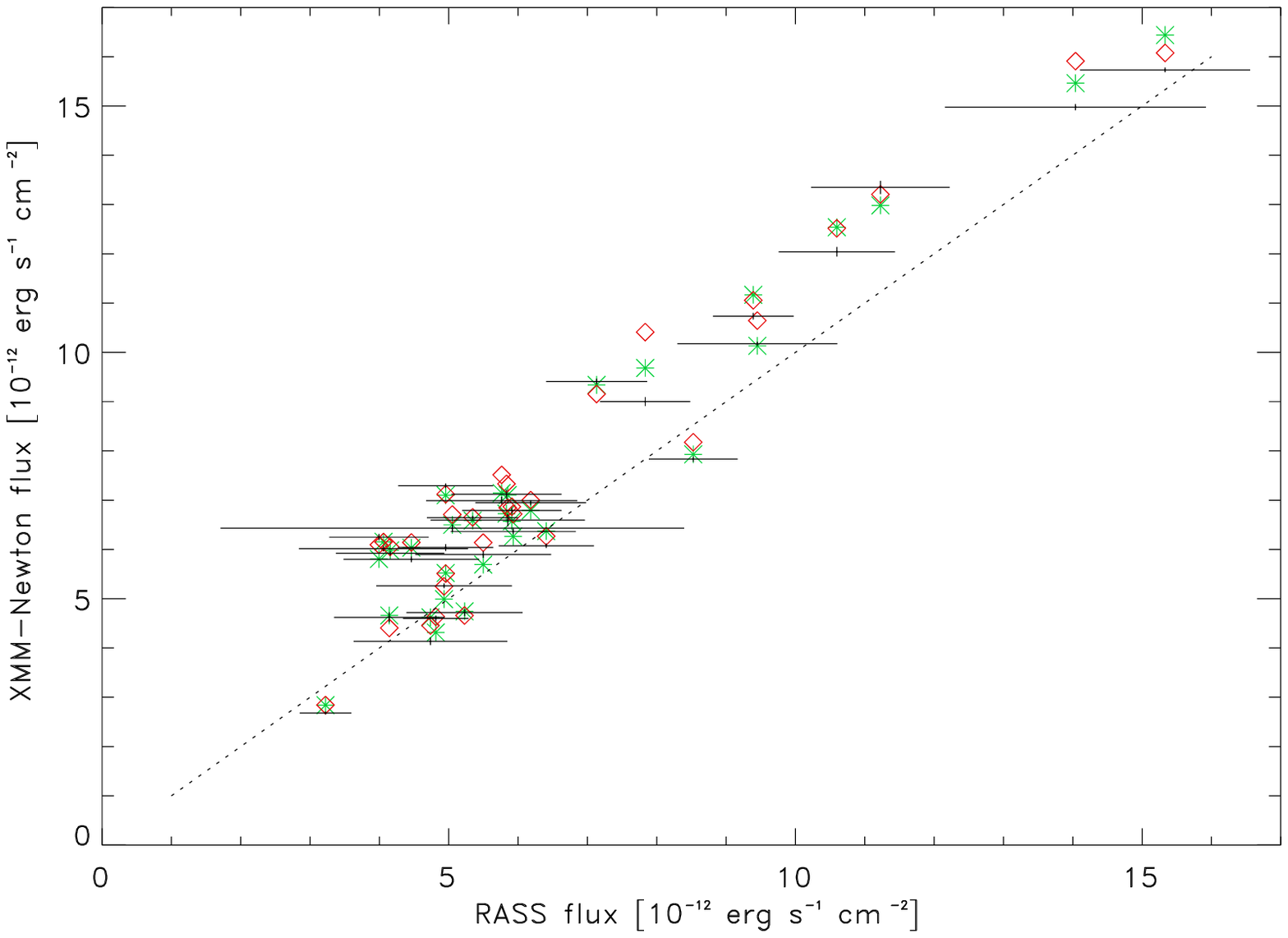} 
   \includegraphics[width=\columnwidth]{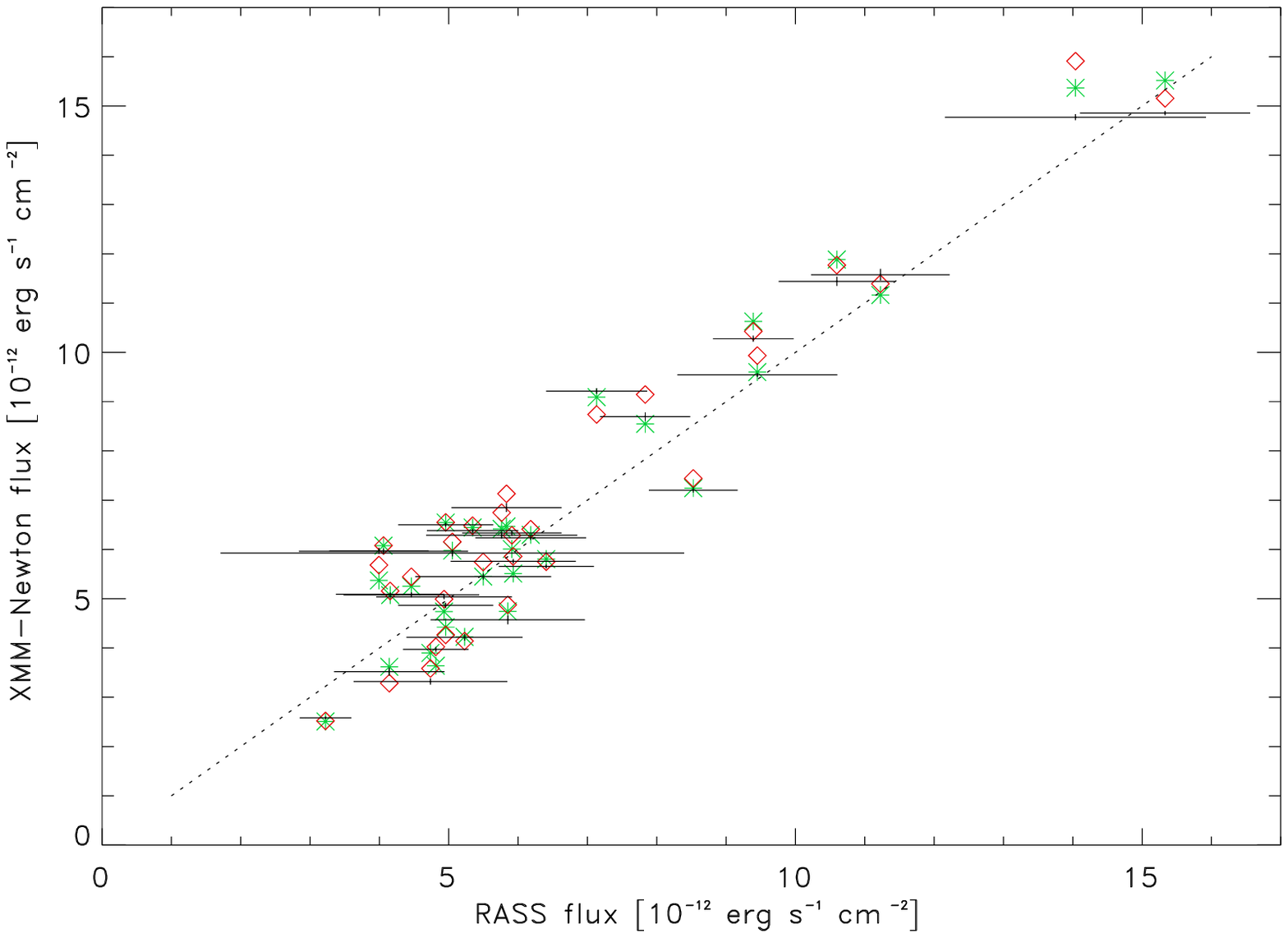} 
      \caption{Comparison of REFLEX clusters fluxes determined from 
      the ROSAT All-Sky Survey and the fluxes obtained from the three 
      detectors of XMM-Newton: black crosses (pn), diamonds 
      (MOS1), green stars (MOS2). The upper panel shows 
      the results for the total XMM-Newton cluster fluxes, while the 
      lower panel shows the XMM-Newton cluster fluxes with point 
      source contamination removed. Error bars are only shown at 
      the location of the pn data points. The vertical error 
      bars are generally smaller than the plotted symbols.} 
         \label{Fig13} 
   \end{figure} 
   \begin{figure} 
   \centering 
   \includegraphics[width=\columnwidth]{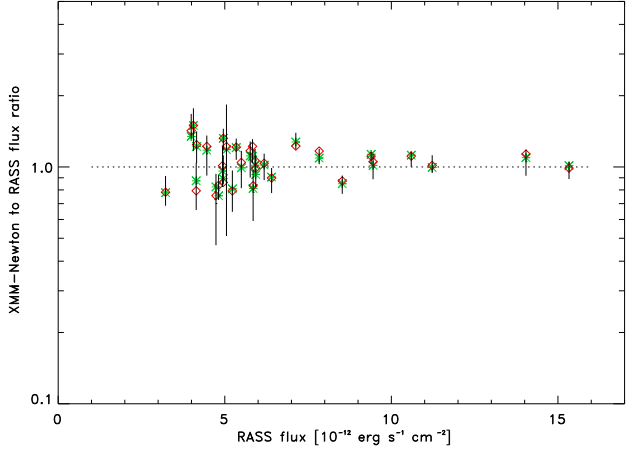} 
      \caption{Ratio of the XMM/Newton 
      observations (contaminating point sources subtracted) to the 
      RASS observations as a function of RASS flux. The symbols have 
      the same meaning as in Fig. 13.} 
         \label{Fig14} 
   \end{figure} 
 
Here we are primarily interested in assessing the reliability with 
which the cluster fluxes have been determined in the RASS 
data. Therefore we take the fluxes determined by B\"ohringer et 
al. (2004), which were obtained with the growth curve analysis method 
for a certain aperture radius (before the correction to total 
fluxes). We apply the growth curve technique to the XMM-Newton data 
out to the same aperture radius, separately for the three detectors 
(since we have sufficiently good statistics), and compare all four results in 
Figs. 13 and 14.  For the XMM-Newton data we can reliably determine 
the growth curve flux only to a maximum radius of 12 arcmin because of 
the XMM-Newton field-of-view, while for 7 clusters in the sample the 
RASS measurement aperture is larger than this (see Tab. 3). Therefore 
we have estimated an upper limit on the extra flux that might be seen 
in the RASS at the larger radii. It is smaller than 2\% for three of 
the clusters (RXCJ0605-3518, RXCJ0645-5413, RXCJ0956-1004), smaller 
than 6\% for three further clusters (RXCJ0457-3152, RXCJ1516-0056, 
RXCJ2157-0747) and larger by $\le 12\%$ for RXCJ0616-4748. This 
is, apart from the last case, smaller than the quoted 1$\sigma$ 
uncertainty.  Note also that the angular resolution of the RASS is 
much worse (more than one arcmin) than that of the XMM-Newton 
observations and therefore the unsharp apertures are not exactly the 
same.  
 
There is a good agreement within the uncertainties of the RASS flux 
determination. In Fig. 13 we compare both the XMM-Newton fluxes and 
the point source subtracted XMM-Newton fluxes, with the RASS data, for 
the same detection aperture.  While the unsubtracted XMM-Newton fluxes 
are on average about 10 \% higher than the RASS fluxes (the 
intercalibration of the two instruments is not known to much better than 
about 5\%), the point source corrected fluxes are in the average only 
about 2\% different. Fig. 14 shows that the deviations in the RASS 
fluxes decreases with the flux level as would be expected. Thus we 
conclude that the REFLEX catalogue contains very reliable flux 
estimates in spite of the very low number of photons available. These 
good results are made possible by the very low X-ray background of the 
RASS. This is also reflected by the fact that with the present 
XMM-Newton data, even with the superb photon statistics, we cannot 
extend the flux measurement to much larger aperture radii than was 
done with the RASS data. 
 
Almost as important as a good flux measurement for the REFLEX 
catalogue are good estimates for the flux uncertainties. The latter 
parameter is also an important input into the construction of a 
precise cosmological model test (e.g. in analogy to Stanek et 
al. 2006). If Fig. 15 we test the reliability of this parameter, where 
we compare the flux uncertainty estimates for the RASS  
results with the deviations between RASS and XMM-Newton fluxes 
(assuming, to first order, that the uncertainties in the XMM-Newton 
fluxes are insignificant). Again we find excellent agreement. 
 
Finally, Fig. 16 provides the statistics of the point source contribution 
to the total cluster flux in the {\sl REXCESS} sample. We have no 
clusters where a point source is dominant. Excluding the complex 
supercluster A901/902, the unsufficient data set of RXCJ0145.0-5300, 
and the AGN RXJ1350.7-3343 from the analysis, we find a mean flux 
contamination of only $\sim 11 \%$, and none of the clusters has a 
larger contamination than 26\% by point sources, as already expected 
from general tests on the REFLEX data. 
 
   \begin{figure} 
   \centering 
   \includegraphics[width=\columnwidth]{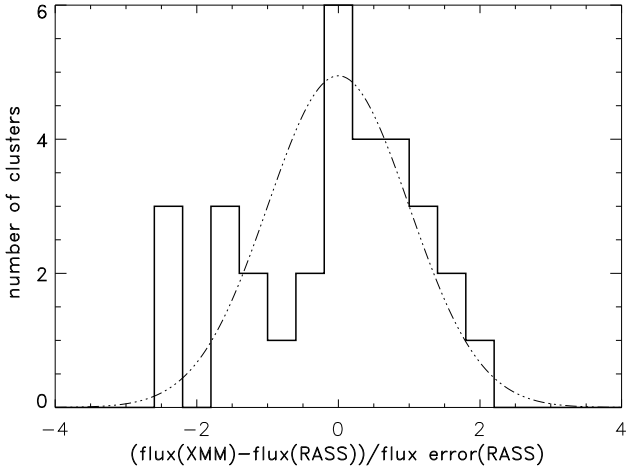} 
      \caption{Ratio of the measured flux difference between 
      XMM-Newton (average of all three detectors) and RASS to the 
      estimated error of the RASS.  Also shown is a Gaussian 
      distribution with $\sigma = 1$, normalized to the total number 
      of clusters. The good agreement of the two curves (except for the 
      three outliers at low XMM-Newton flux) show that the estimated 
      flux errors of the REFLEX sample are precise and reliable. The 
      symbols have the same meaning as in Fig. 13.} 
         \label{Fig15} 
   \end{figure} 
   \begin{figure} 
   \centering 
   \includegraphics[width=\columnwidth]{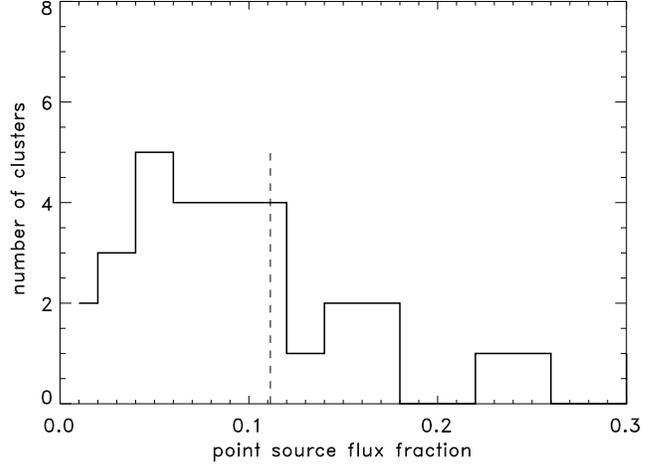} 
      \caption{Statistics of the point source contribution to the 
      cluster fluxes for the sample (excluding 1 complex supercluster, 
      one insufficient data set, and the AGN, RXJ1350.7-3343). The 
      fraction is calculated by dividing the point source contribution 
      by the point source subtracted cluster flux.  The flux 
      contamination is in all cases smaller than 26\% and is on average 
      about 11\%.} 
         \label{Fig16} 
   \end{figure} 
%

\section{Construction of distribution functions} 
 
As the survey volume of the cluster sample is well defined, we can 
construct absolute distribution functions for properties of these 
galaxy clusters. As an example we reconstruct the X-ray luminosity 
function of this sample by means of the cluster densities derived in 
Section 2. Fig. 17 shows the results for the luminosity function using 
both selection schemes outlined in Section 2. The results are compared 
to the REFLEX X-ray luminosity function derived in B\"ohringer et 
al. (2002), which provides the luminosity function for the REFLEX 
sample as observed without an evolution correction.  There is a good 
agreement between the results of the subsample and the total survey 
sample, with the largest deviation in the two highest 
luminosity bins (although these deviations are within the 
errors). This effect is due to the deficiency in the Southern sky of  
luminous X-ray clusters in the nearby Universe. 
The effect is illustrated by the difference of the two selection recipe 
methods: if we use the nearest neighbour boundaries, the last two bins 
extend to $z = 0$, resulting in a smaller cluster density in better 
agreement with the overall REFLEX result. This shows that the difference of 
the density of the most massive clusters in REFLEX and in the 
subsample is due to a real density variation in the Universe. The most 
massive clusters are highly biased and unevenly distributed in the 
REFLEX volume. The general good agreement of the two methods of the 
selection function construction shows that our approach is robust. 
 
In this analysis we have assumed that the uncertainty in the 
luminosity measurement in the {\sl REXCESS} sample is negligible. This 
uncertainty was taken into account in the analysis of the RASS data, 
where the uncertainties are larger (B\"ohringer et al. 2002). 
 
   \begin{figure} 
   \centering 
   \includegraphics[width=\columnwidth]{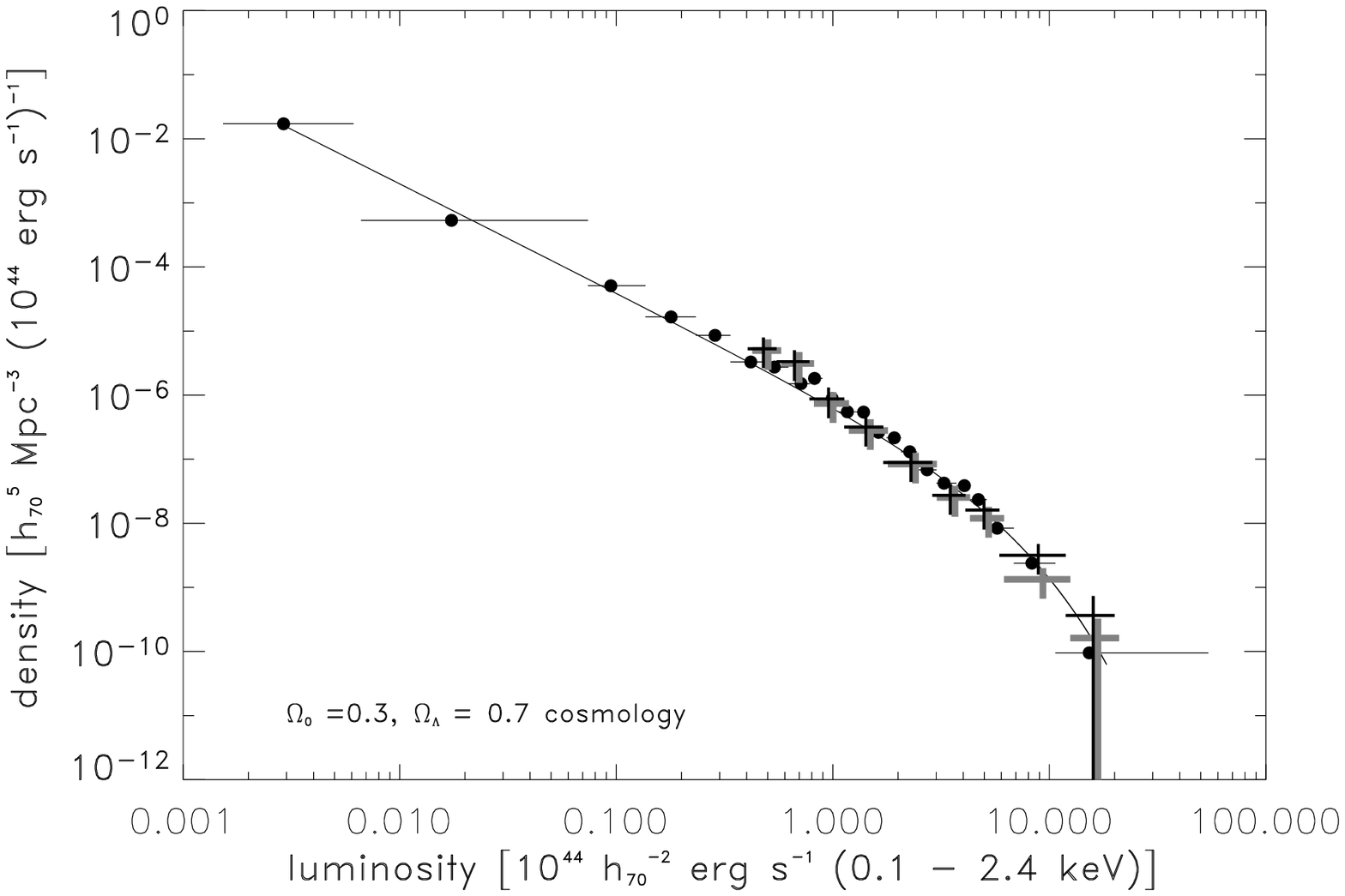} 
   \includegraphics[width=\columnwidth]{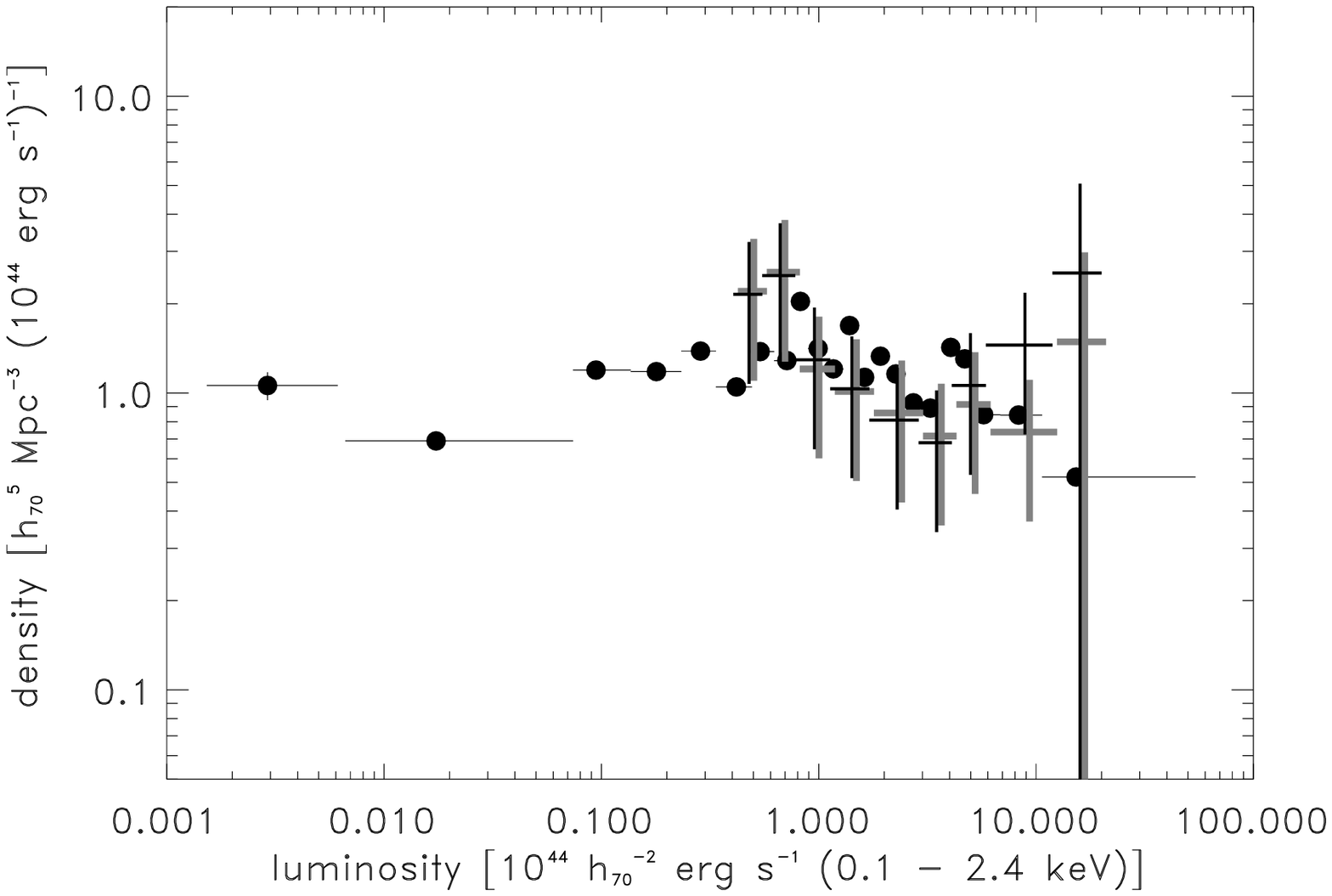} 
      \caption{X-ray luminosity function for the REFLEX sample and for 
      the {\sl REXCESS} subsample; the latter has been constructed by 
      the two different methods explained in the text. Solid dots with 
      error bars refer to the REFLEX function, black crosses mark 
      the function constructed with the original selection function, 
      and grey crosses the function obtained with the alternative 
      method (Section 2). The lower panel shows the same data divided by the 
      Schechter function fit to the REFLEX X-ray luminosity function 
      derived in B\"ohringer et al. (2002).} 
         \label{Fig17} 
   \end{figure} 

\section{Summary and Conclusions} 
 
We have described a sample of 33 galaxy clusters which are selected purely 
on the basis of their X-ray luminosity in nearby redshift shells. The 
sample is therefore representative of an unbiased, X-ray luminosity or 
flux selected subset of the galaxy cluster population.\footnote{A flux 
  limit is an  
effective luminosity selection for each redshift shell.}  The study is 
designed to make the best use of the XMM-Newton observatory to provide 
comprehensive galaxy cluster structure statistics, and representative 
scaling relations. 
 
The results show that the observational results from the REFLEX 
Cluster Survey in the RASS are recovered with excellent agreement, 
except for one REFLEX cluster candidate which was found here to be an 
X-ray AGN. The redetermined X-ray fluxes agree within a few percent 
and the flux errors are also in good agreement. 
 
A detailed description of the cluster sample selection function allows 
us to determine the space density of any subsample with certain 
properties. We demonstrated how the selection function can be applied 
for the evaluation of the distribution function of cluster properties 
for the case of the X-ray luminosity function. The majority of the 
clusters show a roughly regular appearance, very often with 
elongations. Only a few clusters feature several components or peaks 
in the X-ray surface brightness distribution.  Some clusters have a 
somewhat diffuse, low surface brightness structure. There is no 
dramatic merging cluster among the objects in the sample, indicating 
that these systems are probably rare in the general cluster population 
at $z \le 0.2$. 
 
This paper introduces the survey project, for which a series of papers 
covering extensive cluster structure analyses is in progress. 
 
\begin{acknowledgements} We thank our colleague and coauthor Peter
Schuecker, who is sadly no longer with us, for many years of
friendship and fruitful collaboration.

The paper is based on observations obtained with XMM-Newton, an ESA 
science mission with instruments and contributions directly funded by 
ESA Member States and the USA (NASA). The XMM-Newton project is 
supported by the Bundesministerium f\"ur Bildung und Forschung, 
Deutsches Zentrum f\"ur Luft und Raumfahrt (BMBF/DLR), the Max-Planck 
Society and the Haidenhain-Stiftung. GWP acknowledges support from DfG 
Transregio Programme TR33; DP acknowledges support by the German BMBF 
through the Verbundforschung under grant no. 50 OR 0405; AKR 
acknowledges partial financial support from research grants 
MIRG-CT-2004-513676 and NASA LTSA 8390; KP acknowledges support from 
Instrument Center for Danish Astrophysics (The Dark Cosmology Centre 
is funded by the Danish National Research Foundation); HQ thanks the 
FONDAP Centro de Astrofisica for partial support. 
 
\end{acknowledgements}

\end{document}